\pdfoutput=1
\documentclass[fleqn,usenatbib]{mnras}

\usepackage{newtxtext,newtxmath}

\usepackage[T1]{fontenc}
\usepackage{ae,aecompl}
\usepackage[normalem]{ulem}

\usepackage{graphicx}	
\usepackage{float}
\usepackage{amsmath}	
\usepackage{amssymb}	

\newcommand{\oii}{[\ion{O}{II}]}
\newcommand{\oiii}{[\ion{O}{III}]}
\newcommand{\nii}{[\ion{N}{II}]}
\newcommand{\sii}{[\ion{S}{II}]}
\newcommand{\hii}{\ion{H}{II}}
\newcommand{\Rin}{$R_{\mathrm{in}}$}
\newcommand{\Rinp}{$R_{\mathrm{in}}^{\prime}$}
\newcommand{\Rout}{$R_{\mathrm{out}}$}
\newcommand{\RSFR}{$R_{\mathrm{SFR}}$}
\newcommand{\Msun}{M_{\odot}}

\defcitealias{Ma17}{M17}
\defcitealias{Gibson13}{G13}

\title{Gas-phase metallicity gradients of TNG50 star-forming galaxies}

\author[Hemler et al.]{
Z. S. Hemler$^{1}$\thanks{E-mail: zhemler@ufl.edu},
Paul Torrey$^{1}$,
Jia Qi$^{1}$,
Lars Hernquist$^{2}$, 
Mark Vogelsberger$^{3}$, \newauthor
Xiangcheng Ma$^{4}$,
Lisa J. Kewley$^{5,6}$,
Dylan Nelson$^{7}$,
Annalisa Pillepich$^{8}$, \newauthor
R\"{u}diger Pakmor$^{7,9}$,
Federico Marinacci$^{10}$\\
$^{1}$Department of Astronomy, University of Florida, 211 Bryant Space Sciences Center, Gainesville, FL 32611, USA \\
$^{2}$Institute for Theory and Computation, Harvard-Smithsonian Center for Astrophysics, Cambridge, MA 02138, USA \\
$^{3}$Department of Physics, Kavli Institute for Astrophysics and Space Research, Massachusetts Institute of Technology, Cambridge, MA 02139, USA \\
$^{4}$Department of Astronomy, 501 Campbell Hall 3411, University of California, Berkeley, CA 94720-3411, USA \\
$^{5}$Research School for Astronomy \& Astrophysics, Australian National University, Canberra, Australia, 2611 \\
$^{6}$ARC Centre of Excellence for All Sky Astrophysics in 3 Dimensions (ASTRO 3D) \\
$^{7}$Max-Planck-Institut f\"{u}r Astrophysik, Karl-Schwarzschild-Str. 1, D-85748, Garching, Germany \\
$^{8}$Max-Planck-Institut f\"{u}r Astronomie, K\"{o}nigstuhl 17, 69117 Heidelberg, Germany \\
$^{9}$Heidelberger Institut f\"{u}r Theoretische Studien, Schloss-Wolfsbrunnenweg 35, 69118, Heidelberg, Germany \\
$^{10}$Department of Physics \& Astronomy, University of Bologna, via Gobetti 93/2, 40129 Bologna, Italy
}

\date{Accepted XXX. Received YYY; in original form ZZZ}

\pubyear{2020}

\begin{document}
\label{firstpage}
\pagerange{\pageref{firstpage}--\pageref{lastpage}}
\maketitle

\begin{abstract}
\noindent We present the radial gas-phase, mass-weighted metallicity profiles and gradients of the TNG50 star-forming galaxy population measured at redshifts $z=$ 0--3.
We investigate the redshift evolution of gradients and examine relations between gradient steepness and galaxy properties.
We find that TNG50 gradients are predominantly negative at all redshifts, although we observe significant diversity among these negative gradients.
We determine that the gradient steepness of all galaxies increases approximately monotonically with redshift at a roughly constant rate.
This rate does not vary significantly with galaxy mass.
We observe a weak negative correlation between gradient steepness and galaxy stellar mass at redshifts $z\leq2$.
However, when we normalize gradients by a characteristic radius defined by the galactic star formation distribution, we find that these normalized gradients remain invariant with both stellar mass and redshift.
We place our results in the context of previous simulations and show that TNG50 high-redshift gradients are steeper than those of models featuring burstier feedback, which may further highlight high-redshift gradients as important discriminators of galaxy formation models.
We also find that redshift $z=0$ and $z=0.5$ TNG50 gradients are consistent with the gradients observed in galaxies at these redshifts, although the preference for flat gradients observed in redshift $z\gtrsim1$ galaxies is not present in TNG50.
If future JWST and ELT observations validate these flat gradients, it may indicate a need for simulation models to implement more powerful radial gas mixing within the ISM, possibly via turbulence and/or stronger winds.
\end{abstract}

\begin{keywords}
galaxies: ISM -- galaxies: abundances -- galaxies: formation -- galaxies: evolution
\end{keywords}



\section{Introduction}
\label{sec:introduction}




Metallicity is an important physical property of all galaxies. 
Metals are manufactured by aging stellar populations and expelled into the interstellar medium (ISM). 
Subsequently, these metals can be carried along with gas flows~\citep[][]{Lacey85, Friedli94}, diluted by pristine gas accretion \citep[][]{Dekel09, Cresci10}, and/or mixed by mergers \citep[e.g.][]{Kewley10}. 
Given these mechanisms of metal production, redistribution, and dilution, the gas-phase metal abundance of a galaxy retains information regarding the combined galactic history of star formation, gas flows, accretion, and mergers. 
Thus, measurements of metallicity are valuable tools in the study of galaxy formation. 

One key consequence of the integrated co-evolution of galaxies and their metal content is the mass-metallicity relation (MZR), which describes the tight correlation observed between galaxy stellar mass and metallicity \citep[e.g.][]{Tremonti04, Mannucci10, Steidel14, Wuyts14, Sanchez19b}. 
Speculation of a mass-metallicity relation began when early observations of nearby galaxies revealed that galaxy metallicity increases with \textit{B}-band luminosity, a proxy for galaxy mass \citep[e.g.][]{vandenBergh68, Peimbert70, Faber73, Lequeux79}. 
The MZR has been observed to persist out to redshift $z \sim 4$ (\citealt{Finkelstein12} even suggests $z \sim 7-8$), with the normalization of the relation shifting toward lower metallicity values at higher redshifts \citep[e.g.][]{Savaglio05, Erb06, Halliday08, Maiolino08, Hayashi09, Mannucci09, Sanders15, Sanders20, Cullen19}. 
Simulations demonstrate that the MZR is strongly dependent on feedback-driven outflows \citep[e.g.][]{Dave11II, Torrey14, DeRossi17}. 
Reproducing the MZR requires that galaxies drive significant outflows, and that outflow efficiencies are stronger for lower-mass systems \citep[e.g.][]{Brooks07, FD08}. 
These outflows transport metals to the circum- and intergalactic medium (CGM and IGM). 
Enriched CGM gas is then recycled back into the ISM with high efficiency, helping to form the low-mass end of the MZR \citep[e.g.][]{Ma16, Muratov17}. 
As couriers of metals, feedback-driven gas outflows are critical to the MZR and galaxy formation. 

On spatially-resolved scales, the metallicity of galactic disks is observed to decrease with galactocentric distance (e.g. \citealt{Searle71}; \citealt{Smith75}; \citealt{SS78}). 
These negative metallicity gradients can be qualitatively explained via inside-out disk growth models wherein stellar mass initially builds up at the galactic center before forming at progressively greater distances \citep[e.g.][]{MF89, BP99}. 
The negative stellar mass gradient produces a corresponding negative metallicity gradient as aging stellar populations return metals to the ISM \citep[e.g.][]{Ho15}. 
Indeed, this gradient-produces-gradient behavior should exist for any disk growth model if enriched gas remains undisturbed. 
However, the same feedback-driven outflows required to shape the MZR can also act to flatten metallicity gradients~\citep[e.g.][]{Gibson13, Marinacci14}. 
More generally, any process that radially redistributes gas (e.g. outflows and turbulence) will homogenize gas metallicity and level metallicity gradients. 
Additionally, dilution of enriched gas via accretion of pristine IGM gas onto the galactic disk can act to flatten or possibly even invert metallicity gradients, as can galactic fountains or mergers \citep[e.g.][]{Cresci10, Kewley10, Rupke10a, Brook12a, Fu13, Troncoso14}. 
As products of these many phenomena, metallicity gradients are rich but complicated indicators of galaxy assembly history, gas dynamics, feedback, and accretion. 

For the Milky Way and local galaxies, gas-phase metal abundances can be measured by inferring electron temperatures from auroral lines of planetary nebulae \citep[e.g.][]{Maciel03} or \hii\, regions \citep[e.g.][]{Stanghellini14}. 
This method (hereafter the ``electron temperature method'') is outlined in \citet{Perez-Montero17}. 
However, these auroral lines are too weak to be easily observed in distant galaxies via the current generation of telescopes, although recent works \citep[e.g.][]{Andrews13, Izotov15, Ly16, Yates20} have made global metallicity measurements using the electron temperature method for galaxies at redshifts $z \lesssim 1$. 
Other abundance measurement methodologies (e.g. recombination lines, stellar absorption lines) are similarly limited \citep[][]{Yates20}. 
Instead, most spatially-resolved extragalactic metallicities (and therefore gradients) are measured via relations between metal abundances and strong nebular emission lines associated with star-forming \hii\, regions \citep{Kewley19}. 
At first, these strong-line relations were determined empirically and calibrated using metallicity measurements from auroral lines \citep[e.g.][]{Alloin79, PES80}. 
Later, more solidified calibrations came via further auroral line measurements \citep[e.g.][]{Pilyugin01, PP04} and theoretical models of \hii\, regions \citep[e.g.][]{Dopita86, Dopita00, Kewley02, Dopita13}. 
Strong lines commonly utilized for these measurements are \oii, \oiii, \nii, and \sii, which are sufficiently bright to be observed out to significant cosmic distances. 
Specifically, the ratios of these strong lines, termed ``indicators'' (e.g. $\mathrm{N2} \equiv \log \frac{\nii \lambda 6583}{\mathrm{H}\alpha}$), are used to evaluate ``calibrators'' (e.g. PP04-N2 -- \citealt{PP04}) that return abundance inferences. 
In this manner, oxygen abundance gradients have been measured for a large range of systems, including targeted isolated local galaxies \citep[e.g.][]{Zaritsky94, vanZee98}, local interacting galaxies \citep[e.g.][]{Kewley10, Rupke10b}, large surveys of local galaxies  \citep[e.g.][]{Sanchez12b, Sanchez14, Sanchez19b, Sanchez-Menguiano16, Sanchez-Menguiano18, Belfiore17, Poetrodjojo18}, and out to high redshift in unlensed \citep[e.g.][]{Swinbank12, Wuyts16, Molina17, ForsterSchreiber18} and lensed \citep[e.g.][]{Stark08, Yuan11, Jones15, Leethochawalit16, Wang19, Curti20b} systems. 

This paper takes a particular interest in the gas-phase metallicity gradients of high-redshift galaxies for reasons that will be detailed shortly. 
First, we will briefly summarize the history and current state of high-redshift gradient measurements. 
The earliest measurements of these gradients \citep[][]{Stark08, Jones10, Yuan11, Jones13}, now roughly a decade old, came via adaptive optics-assisted observations of gravitationally-lensed galaxies. 
These observations were made using the Keck II telescope and its OSIRIS integral field spectrograph \citep[][]{Larkin06}. 
While \citet{Stark08} measured a flat gradient in its unrelaxed redshift $z \sim 3$ galaxy, \citet{Jones10}, \citet{Yuan11}, and \citet{Jones13} all measured gradients much steeper than those observed locally, indicating that gradient steepness increases with redshift. 
Around this time, \citet{Cresci10} and \citet{Swinbank12} made adaptive optics-assisted observations of unlensed galaxies (at redshifts $z \sim 3$ and $z \sim 1.5$, respectively) using the ESO VLT and its SINFONI integral field spectrograph \citep[][]{Eisenhauer03}. 
\citet{Cresci10} measured inverted gradients, while \citet{Swinbank12} measured negative gradients that were as shallow -- and in some cases, more shallow -- than those observed at lower redshifts. 
In subsequent years, there came many more measurements of high-redshift gradients in both lensed and unlensed galaxies via observations that either were or were not seeing-limited \citep[][]{Troncoso14, Jones15, Leethochawalit16, Wuyts16, Molina17, Wang17, ForsterSchreiber18, Wang19, Curti20b}. 
In spite of the significant progress that resulted from these observations, there still seems to exist some level of tension between them. 
As with \citet{Jones10}, \citet{Yuan11}, and \citet{Jones13} vs. \citet{Swinbank12} vs. \citet{Cresci10}, several studies continue to measure some steep negative gradients, while others measure only gradients either consistent with or flatter than those of the local Universe, and still others measure significantly inverted gradients (see Section~\ref{sec:dtension}). 
Thus, the precise distribution of high-redshift gradients currently does not appear fully constrained. 

Even in the local Universe, there is not necessarily broad agreement among gradient surveys -- for example, CALIFA and AMUSING \citep[][]{Sanchez12b, Sanchez14, Sanchez-Menguiano16, Sanchez-Menguiano18} find significant evidence for a characteristic effective radius-normalized metallicity profile/gradient among local galaxies, while MaNGA \citep[][]{Belfiore17} does not. 
Moreover, MaNGA and SAMI \citep[][]{Poetrodjojo18} observe positive correlations between galaxy stellar mass and normalized gradient steepness. 
CALIFA and AMUSING, however, do not. 
See \citet{Sanchez19a} for a detailed review of low-redshift metallicity gradient survey discrepancies. 

Several potential sources of systematic error in gradient measurements could be contributing to discrepancies between studies. 
Many previous works \citep[e.g.][]{Yuan13, Mast14, Poetrodjojo19, Acharyya20} have found that observations with low angular resolution and signal-to-noise ratios can cause gradients to appear significantly flattened. 
Some studies propose that this could be a result of beam smearing and the apparent overlap of \hii\, regions with regions of diffuse ionize gas (DIG) and/or other \hii\, regions with differing physical properties \citep[e.g.][]{Yuan13, Zhang17, Poetrodjojo19, Kewley19}. 
Moreover, low spectral resolution could lead to inaccurate abundance measurements, as shocked gas contributes a non-thermal component to strong lines that must be removed -- otherwise, the assumptions of theoretical metallicity diagnostic models are violated \citep[][]{Rich11, Kewley13}. 
Additionally, the models and assumptions of strong-line calibrations themselves may be inaccurate, especially at higher redshifts \citep[][]{Steidel16, Strom17, Carton18, Kewley19}. 
It is also well-known that abundances derived using differing calibrators often disagree significantly -- even those that use the same indicators. 
Some calibrators are based on electron temperature observations in local \hii\, regions (e.g. O3N2-M13 -- \citealt{Marino13}), some on photonionization models (e.g. \textsc{pyqz} -- \citealt{Dopita13}), and some on both (e.g. PP04-O3N2 -- \citealt{PP04}) \citep[][]{Sanchez19b}. 
Additionally, some indicators are not equally suited for inferring metallicity in all environments --  see \citet{Maiolino19} for an overview of the strengths and weaknesses of each indicator. 
Because different studies often use different indicators/calibrators, the resulting disagreements complicate direct comparisons between studies and with theory, although deviations can be lessened via appropriate conversions \citep[e.g.][]{Kewley08}. 
Still, this calibrator-dependence may extend to gradients, bringing into question the accuracy of strong-line gradient measurements. 
See Section~\ref{sec:dtension} and \citet{Kewley19} for a more robust discussion of strong-line calibrations and their potential systematics. 

\begin{figure}
\centerline{\vbox{\hbox{
\includegraphics[width=0.45\textwidth]{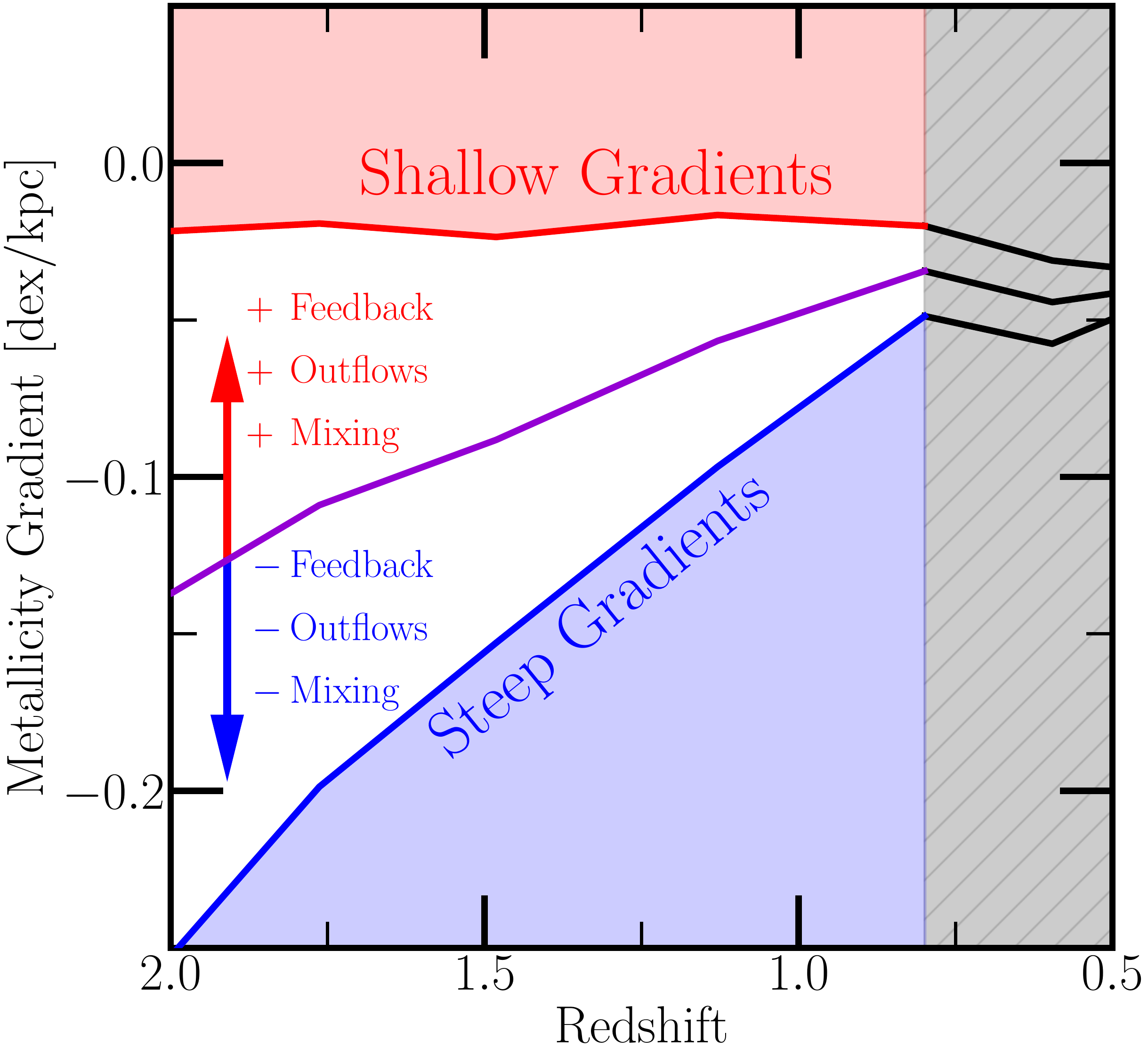} 
}}}
\caption{
A schematic summarizing the results of \citet{Gibson13}. 
We illustrate how adjusting the strength of feedback models affects the redshift evolution of metallicity gradients. 
The gradient evolution of enhanced (i.e., strong) and conventional (i.e., weak) feedback models are shown in red and blue, respectively. 
The gradient evolution of a hypothetical model with feedback stronger than conventional models but weaker than enhanced models is shown in purple. 
One can imagine adjusting the purple evolutionary track along the red (blue) arrow by strengthening (weakening) its feedback model. 
The evolution of conventional feedback models diverges from that of enhanced feedback models around redshift $z \sim 1$. 
By $z \sim 2$, the gradients predicted by conventional and enhanced feedback models are drastically different. 
Thus, if observers measure steep (shallow) gradients at $z \sim 2$, this may imply that our Universe favors a conventional (enhanced) feedback model.
}
\label{fig:schm}
\end{figure}

High-redshift metallicity gradients are particularly interesting because, as galaxy simulations \citep[][]{Pilkington12b, Gibson13} have revealed, they are sensitive to differing feedback implementations. 
Figure~\ref{fig:schm} displays a schematic, based on the results of \citet{Gibson13} (hereafter \citetalias{Gibson13}), that illustrates how the redshift evolution of gradients responds to feedback. 
Specifically, \citetalias{Gibson13} found that a model featuring ``conventional'' feedback \citep[i.e., MUGS;][]{Stinson10, Pilkington12b} leads to steep gradients at high redshift, while a model featuring ``enhanced'' feedback \citep[i.e., MaGICC;][]{Brook11, Brook12a, Brook12b, Brook12c, Pilkington12a, Stinson12,Stinson13} leads to shallow gradients at high redshift.\footnote{
\citetalias{Gibson13} defines conventional feedback models as those which inject $\sim 10-40\%$ of supernova (SN) energy into the ISM as heat -- for example, MUGS injects $4 \times 10^{51}$ erg/SN. 
On the other hand, MaGICC -- the enhanced feedback model -- injects $10^{51}$ erg/SN and includes radiation energy feedback.
}
Gradients produced by conventional feedback models like MUGS steepen with redshift (and flatten with time) because weaker feedback facilitates inside-out galaxy formation. 
Galaxies formed in this manner begin with centrally-concentrated star formation that enriches only the inner disk, giving rise to steep gradients. 
These steep gradients gradually flatten with time as star formation shifts to progressively larger radii \citep[e.g.][]{PB00}. 
On the other hand, enhanced feedback models like MaGICC (i) suppress concentrated star formation and (ii) efficiently redistribute enriched gas, thereby producing shallow gradients at all times. 
This divergent gradient redshift evolution suggests that high-redshift gradients can be used to constrain galaxy formation models. 
High-redshift metallicity gradients will therefore be a topic of interest in the coming era of JWST and the ELTs -- especially because these telescopes and their instruments will be capable of overcoming all the aforementioned potential gradient measurement systematics \citep[][see Section~\ref{sec:dtension}]{Yuan11, Wuyts16, Wang19, Maiolino19, Curti20b}. 
Thus, in advance of this coming era, it is important to have clear predictions for these gradients. 

This paper presents the gas-phase metallicity profiles and gradients of the TNG50 \citep[][]{Nelson19b, Pillepich19} simulated galaxy population. 
We investigate relations between gradient steepness and galaxy properties including stellar mass, size, and kinematics. 
Moreover, we analyze the redshift evolution of these relations. 
We compare our results to the gradients of observed and simulated galaxies and make predictions for the gradients of galaxies out to redshift $z = 3$. 
While the global metallicities of TNG100 galaxies have already been studied and found to be consistent with MZR observations out to redshift $z = 2$ \citep[][]{Torrey19}, a spatially-resolved metallicity analysis requires the higher resolutions now afforded by TNG50. 
Other recent works have analyzed metallicity gradients for a small sample of galaxies out to high redshift using zoom-in simulations \citep[FIRE;][]{Ma17} and for a large sample of galaxies at redshift $z = 0$ using a cosmological simulation \citep[EAGLE;][]{Tissera19}. 
This work combines advantages of \citet{Ma17} and \citet{Tissera19}, being the first metallicity gradient analysis of a large simulated galaxy sample ($>$6000) in full cosmological context from redshift $z = 0$ out to $z = 3$. 

The structure of this paper is as follows. 
In Section~\ref{sec:methods} we describe the TNG50 simulation and dataset, our galaxy selection criteria, and our methods for measuring the physical properties, metallicity profiles, and metallicity gradients of these galaxies. 
In Section~\ref{sec:results} we present our main results, including the redshift evolution of metallicity gradients and relations between these gradients and galaxy physical properties. 
Additionally, we place our results in the context of gradient measurements from previous studies.
In Section~\ref{sec:discussion} we discuss our results and how they relate to these previous simulations and observations. 
Finally, in Section~\ref{sec:conclusion} we summarize our findings and state our conclusions. 

\section{Methods}
\label{sec:methods}

In this paper, we present an analysis of the gas-phase metallicity gradients of star-forming galaxies in the TNG50 dataset. 
This section gives a brief overview of the IllustrisTNG simulation suite, our galaxy selection criteria, and our methods for measuring galaxy physical properties, kinematics, metallicity profiles, and metallicity gradients. 
Much of our methodology closely follows that of \citet[][hereafter \citetalias{Ma17}]{Ma17}. 
For the entirety of our analysis, we measure length in physical units. 

\begin{figure*}
\centerline{\vbox{\hbox{
\includegraphics[width=1.0\textwidth]{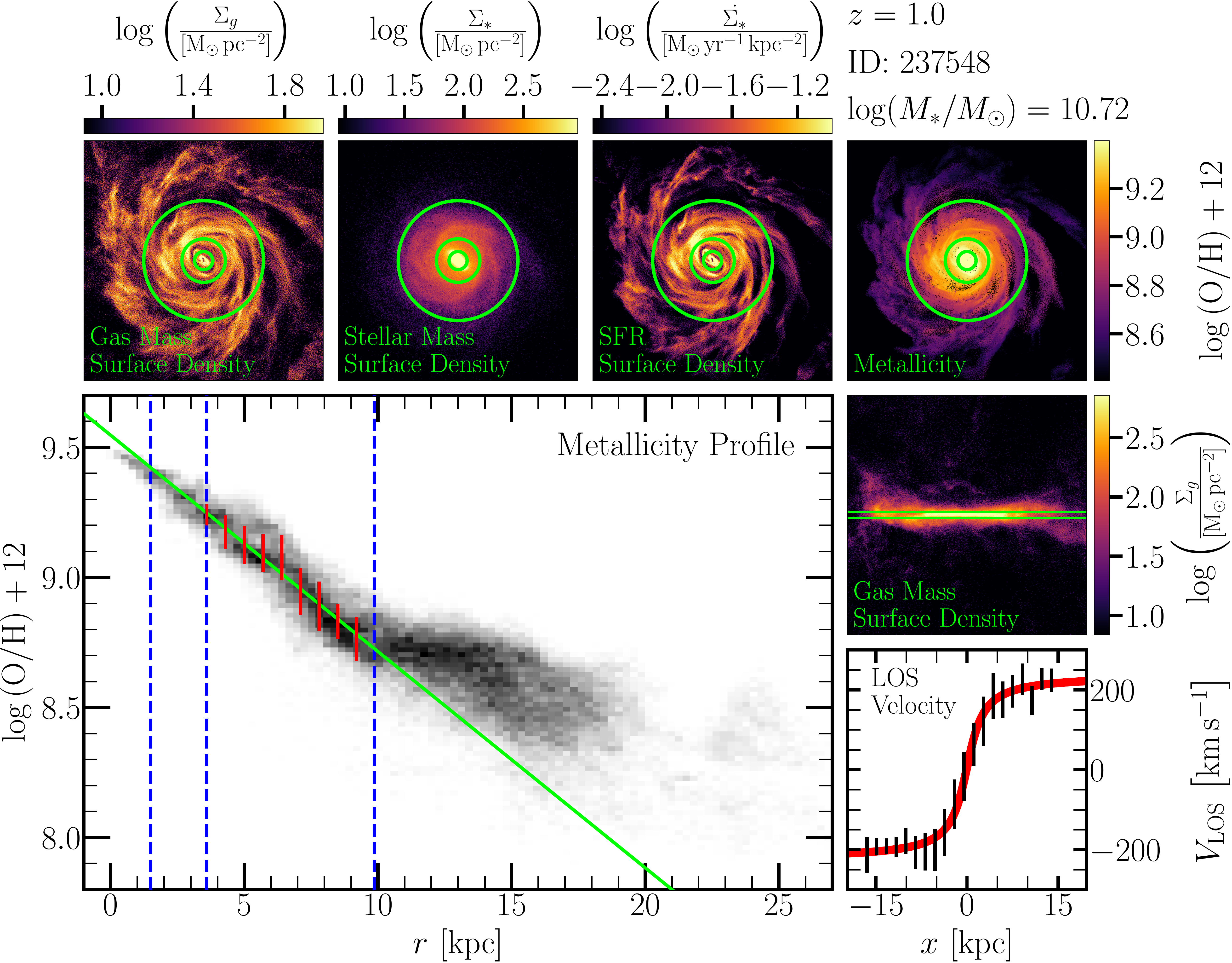}
}}}
\caption{
The maps, gas-phase metallicity profile, and gas-phase metallicity gradient of one galaxy in the TNG50 sample. 
Moving counter-clockwise from the large bottom-left panel, we show the gas-phase mass-weighted metallicity profile and the unweighted line-of-sight gas velocity curve, then maps of edge-on gas mass surface density, face-on gas-phase mass-weighted metallicity, face-on star formation rate surface density, face-on stellar mass surface density, and face-on gas mass surface density. 
For every panel, we measure length in physical units. 
Each map pixel has a width of 100 pc. 
The gas-phase metallicity profile (generated via the methods described in Section~\ref{sec:profgrad}) is plotted as a 2D histogram, with more populated bins displaying a darker shade. 
The three dashed blue lines dividing the metallicity profile denote \Rin, \Rinp, and \Rout \, (see Section~\ref{sec:galphysprops}). 
These radii are also marked by green circles in the galaxy maps. 
The red lines atop the metallicity profile between \Rinp and \Rout \, (i.e., the reduced metallicity profile within the gradient fit region, see Sections~\ref{sec:galphysprops}~and~\ref{sec:profgrad}) mark the median and $1\sigma$ spread of metallicity measured along the metallicity profile. 
The solid green line shows the fit of Equation~(\ref{eq:gradfit}) through these metallicity medians, obtained via the methods described in Section~\ref{sec:profgrad}. 
Figure~\ref{fig:proffit} gives the parameter distributions of this fit. 
In the line-of-sight gas velocity curve, black lines mark the median and $1\sigma$ spread of $V_{\mathrm{LOS}}$ measured along the slit formed by the green horizontal lines (placed 0.5 kpc above and below the galactic plane) in the edge-on gas mass surface density map. 
The red curve shows the least-squares fit of Equation~(\ref{eq:diskfit}) to these $V_{\mathrm{LOS}}$ medians. 
Our methods of measuring galaxy kinematics are fully detailed in Section~\ref{sec:galkinematics}.
}
\label{fig:subs}
\end{figure*}

\subsection{The IllustrisTNG and TNG50 simulations}
\label{sec:tng}

The IllustrisTNG (TNG) cosmological simulation suite \citep[][]{Marinacci18, Naiman18, Springel18, Pillepich18a, Pillepich18b, Pillepich19, Nelson18, Nelson19a, Nelson19b} is the successor of the Illustris simulation \citep[][]{Vogelsberger13, Vogelsberger14a, Vogelsberger14b, Genel14, Torrey14, Sijacki15}. 
TNG, which runs on the moving-mesh code \textsc{arepo} \citep[][]{Springel10}, includes a further refined physical model than that of Illustris. 
Because the physical model and methods of TNG have been detailed elsewhere \citep[][]{Weinberger17, Pillepich18a}, we give only a brief overview of these topics with an emphasis on the simulation elements that directly influence the analysis and results presented in this paper. 

In addition to including self-gravity and (magneto-)~hydrodynamics, the TNG simulations model radiative gas cooling, star formation in dense gas, and feedback from stars and supermassive black holes (SMBHs). 
The TNG model -- including the pressurization of the interstellar medium (ISM), stellar wind properties, and black hole feedback -- was calibrated to broadly recover the observed cosmic star formation rate density redshift evolution and several characteristics of galaxies at redshift $z = 0$, such as the galaxy stellar mass function, black hole mass-stellar mass relation, stellar mass-halo mass relation, and the mass-size relation. 
Three flagship runs constitute the TNG simulation suite, each with a differing cosmological volume and resolution -- $51.7^3 \, \mathrm{Mpc^3}$ with $2 \times 2160^3$ resolution elements (TNG50), $110.7^3 \, \mathrm{Mpc^3}$ with  $2 \times 1820^3$ (TNG100), and $302.6^3 \, \mathrm{Mpc^3}$ with $2 \times 2500^3$ (TNG300). 
This paper analyzes TNG50, which is highest-resolution run in the suite (with star-forming gas cell sizes of $\sim$100 pc on average). 

Critical for the purposes of this paper, the TNG simulations model the return of mass and metals back to the local ISM from aging stellar populations. 
In the simulation, ``star particles'' are stochastically formed whenever and wherever gas is sufficiently dense ($n_{\mathrm{H}} \gtrsim 0.13 \, \mathrm{cm^{-3}}$), representing the birth of unresolved stellar populations. 
These unresolved stellar populations are assumed to follow a~\citet{Chabrier03} initial mass function. 
Newborn stars inherit the metallicity of the local ISM that they originate from. 
As time progresses and stars move off the main sequence, mass and metals are injected back into the ISM surrounding the star particle, increasing the metal abundance of the local ISM. 
For this enrichment process, stellar lifetimes are adopted from \cite{Portinari98}, while mass return rates and metal yields are adopted from \citet{Nomoto97} for SNIa, \citet{Portinari98} and \citet{Kobayashi06} for SNII, and \citet{Karakas10}, \cite{Doherty14}, and \citet{Fishlock14} for AGB stars. 
The abundances of ten different metal species are tracked -- hydrogen, helium, oxygen, nitrogen, magnesium, silicon, sulfur, neon, iron, and europium. 
To facilitate comparisons with observations, we report gas-phase metallicity as the logarithmic abundance ratio of oxygen to hydrogen (see Equation~\ref{eq:abrat}). 

Following injection, metals can be spatially redistributed via many physical processes, thereby altering metallicity gradients. 
While virtually any aspect of the TNG model could influence this metal redistribution, we suspect the model's pressurized equation of state in the ISM and its implementation of galactic winds may both have a significant effect. 
TNG is unable to explicitly model some of the physics that work to pressurize the ISM (e.g. small-scale turbulence, thermal instability/conduction, molecular cloud formation/evaporation; \citealt{Vogelsberger13}) because these phenomena occur on scales beyond the resolution limits of the simulation. 
Instead, TNG models ISM pressurization via a two-phase, effective equation of state model described in \citet{Springel03}. 
However, larger-scale processes like galactic winds are explicitly modeled. 
TNG winds are generated by both star-forming gas and SMBHs. 
Star-forming gas launches stellar winds with velocities proportional to the local velocity dispersion of dark matter. 
These stellar winds are injected isotropically with a mass loading factor derived from the wind speed and available SN wind energy, and a metal loading factor assumed to be some constant fraction. 
Stellar wind mass and metals are carried away from the ISM in which the wind originates. 
Initially, these stellar winds are hydrodynamically decoupled from the local ISM. 
Upon reaching more diffuse regions of the ISM/CGM, stellar winds recouple and deposit their constituent mass, momentum, energy, and metal content. 
SMBHs in both high-accretion and low-accretion states also release feedback energy into surrounding gas, which can drive highly directional, galactic-scale winds (as analyzed in TNG50; \citealt{Nelson19b}). 
This process is particularly effective for low-accretion rate SMBHs, for which the TNG model injects momentum in discrete events, each time in a random direction. 
For high-accretion rate SMBHs, the feedback energy injection is thermal and continuous in time. 
Unlike stellar winds, winds from SMBHs are never hydrodynamically decoupled. 
As with other aspects of the TNG model, wind velocities and energies were calibrated to roughly reproduce the aforementioned redshift $z = 0$ galaxy population characteristics and the cosmic star formation rate density redshift evolution. 
Additionally, the TNG wind velocity implementation was chosen to capture the redshift evolution of galaxy stellar mass and luminosity functions. 
To this end, TNG50 includes redshift-dependent wind velocities that remain invariant with the growth of virial halo mass per \citet{Henriques13}. 
TNG50 also assigns a minimum wind velocity to avoid unrealistically massive winds in low-mass galaxies. 
Moreover, TNG50 assumes that some fraction of wind energy is thermal and that wind energy decreases with increasing metallicity of the originating star-forming gas. 
A full description of and justification for these aspects of the TNG model is given in \citet{Pillepich18a}. 

\subsection{Galaxy selection}
\label{sec:galselection}

Galaxies are identified from the simulation output using the \textsc{subfind} algorithm~\citep[][]{Springel01, Dolag09}, which identifies self-bound collections of particles from within parent Friends-of-Friends groups~\citep[][]{Davis85}. 
Throughout this paper, we limit our analyses to systems where gas-phase metallicity gradients can be reliably resolved. 
While there is no clear-cut divide between well- and poorly-resolved gradients, we find that galaxies with at least $\sim 10^4$ gas resolution elements generally have sufficient radial sampling for robustly determining gradients. 
The highest resolution ($n = 2 \times 2160^3$; TNG50-1) run of TNG50 has a target baryon resolution element mass of $m_{\mathrm{b}} \approx 8.5 \times 10^4 M_{\odot}$, which leads us to only consider systems with gas masses of $M_{\mathrm{gas}} \geq 10^9 M_{\odot}$. 
Additionally, we impose stellar mass cuts on our galaxy sample of $M_{*} \geq 10^{9} M_{\odot}$ and $M_{*} \leq 10^{11} M_{\odot}$. 
These cuts ensure that all mass bins contain well-resolved galaxies, and that each mass bin has a sufficient number of simulated systems for meaningful statistical analyses. 

Gas-phase metal abundances are probed observationally through the strong emission lines of star-forming \hii\, regions \citep[][]{Kewley19}. 
We therefore include only star-forming galaxies in our sample to facilitate an even-handed comparison with observations. 
Following \citet{Donnari19} and \citet{Pillepich19}, we define star-forming galaxies as those with integrated sSFR greater than or equal to the specific star formation main sequence (sSFMS) at their stellar mass, or less than 0.5 dex below the sSFMS at their stellar mass. 
At a given redshift, we construct the sSFMS from median sSFRs in stellar mass bins of 0.2 dex for $M_* \leq 10^{10.2} \, M_{\odot}$. 
For $M_* > 10^{10.2} \, M_{\odot}$, we take the sSFMS to be the least-squares linear fit through the $M_* \leq 10^{10.2} \, M_{\odot}$ medians. 
Finally, we exclude satellite galaxies from our sample. 
We include all self-bound particles of each selected galaxy in our analyses unless otherwise noted.
Maps of stellar/gas mass surface density, SFR surface density, and gas-phase metallicity of a typical high-mass galaxy in our sample are shown in Figure~\ref{fig:subs}. 

\subsection{Galaxy radii and inclination}
\label{sec:galphysprops}

We define the center of each galaxy as the location of its potential minimum. 
Before measuring metallicity profiles and gradients, we first compute galaxy inclination angles and characteristic radii. 
Similar to \citetalias{Ma17}, we define the size of our galaxies using their radial distribution of star formation. 
Specifically, we define two characteristic radii -- \Rin\, and \Rout. \Rin\, is the 3D distance from the center of the galaxy for which $5\%$ of the total SFR of a galaxy is enclosed, while \Rout\, is the distance for which $90\%$ of the total SFR within $10$ kpc of \Rin\, is enclosed. 
We define \Rout\, in terms of \Rin\, because a significant fraction of the galaxies in our late-time sample have extended central cavities void of gas as a result of AGN feedback. 
Hereafter, we refer to the region between \Rin\, and \Rout\, as the ``star-forming region''. 
We also measure \RSFR, the 3D distance from the center of the galaxy for which half the total SFR of a galaxy is enclosed. 
 
We calculate galaxy inclination angles from the vector normal to the galactic mid-plane. 
We define this normal vector to be parallel to the summed angular momenta of all gas cells in the star-forming region of each galaxy. 
These inclination angles are used to rotate galaxies to the edge-on position for the kinematic measurements of Section~\ref{sec:galkinematics} and to the face-on position for the metallicity profile and gradient measurements of Section~\ref{sec:profgrad}. 
A galaxy rotated to both face-on and edge-on orientation via the these inclination angles (with \Rin\, and \Rout\, marked) is shown in Figure~\ref{fig:subs}. 

\subsection{Galaxy kinematics}
\label{sec:galkinematics}

Following \citetalias{Ma17}, we employ a long-slit spectroscopy technique (also similar to that of \citealt{Pillepich19}) to measure the kinematic properties of our galaxies and determine the degree to which they are rotationally supported. 
We begin by rotating a galaxy to the edge-on position (via the inclination angles described in Section~\ref{sec:galphysprops}) and defining a coordinate axis such that the origin is at the galaxy center, the $x$-axis is parallel to the galaxy edge, the $y$-axis is parallel to the line-of-sight, and the $z$-axis is normal to the galaxy mid-plane. 
We exclude all gas more than $0.5$ kpc above and below the galactic mid-plane (i.e., $|z| > 0.5$ kpc). 
These $z = \pm 0.5$ kpc bounds are displayed in the edge-on (i.e., $xy$-plane) galaxy maps of Figure~\ref{fig:subs}, forming the titular long slit of this technique. 
Moreover, we exclude gas cells with hydrogen number density $n_{\mathrm H} < 0.13\,\mathrm{cm^{-3}}$ (the approximate TNG50 star formation density threshold) to avoid contributions from diffuse gas outside the galactic disk. 
If a gas cell meets this density requirement, it is included in our analysis -- we do not explicitly impose an additional star formation requirement. 
In the range $-R_{\mathrm{out}} < x < R_{\mathrm{out}}$ we measure the unweighted median line-of-sight velocity (i.e., $V_y$ or $V_{\mathrm{LOS}}$) and the unweighted $1\sigma$ line-of-sight velocity standard deviation ($\sigma_{V_{\mathrm{LOS}}}$) of the gas cells in each bin with a spatial resolution of $x = 0.1$ kpc. 
We exclude bins containing less than 16 gas cells to avoid unreliable velocity measurements with artificially small uncertainties. 

Like \citetalias{Ma17} and many observational works \citep[e.g.][]{Jones10, Swinbank12, Leethochawalit16}, we fit our line-of-sight velocity curves with the simple disk model
\begin{equation}
\label{eq:diskfit}
    V(r) = V_0 + V_c \left[ \frac{2}{\pi} \arctan{\left( \frac{r}{R_t} \right)} \right]
\end{equation}
which has three free parameters: $V_0$, $V_c$, and $R_t$. 
Because our galaxies always have some non-zero bulk line-of-sight velocity, the fit parameter $V_0$ is required as an additive normalization that shifts the fit into the galaxy line-of-sight rest-frame. 
Respectively, $R_t$ and $V_c$ scale the width and amplitude of the $\arctan$ function such that the fit asymptotes to $V_0 \pm V_c$ for $r \gg R_t$. 
This functional behavior is physically motivated by the rotational velocity of a well-ordered disk asymptoting to some maximal velocity at large galactocentric distances. 
We require that $V_0 + V_c \leq V_{\mathrm{max}}$ and $V_0 - V_c \geq V_{\mathrm{min}}$, where $V_{\mathrm{max}}$ and $V_{\mathrm{min}}$ are respectively the maximum and minimum $V_{\mathrm{LOS}} + \sigma_{V_{\mathrm{LOS}}}$ and $V_{\mathrm{LOS}} - \sigma_{V_{\mathrm{LOS}}}$ along the velocity curve. 
Moreover, we require that $V_c$ > 0 and that $0 < R_t < 2R_{\mathrm{out}}$. 
We fit Equation~(\ref{eq:diskfit}) only to the region $-R_{\mathrm{out}} \leq x \leq R_{\mathrm{out}}$. 
If a discontinuity in the velocity curve greater than 1 kpc exists at some $x^{+}_{\mathrm{dis}}$ where $R^{\prime}_{\mathrm{in}} < x^{+}_{\mathrm{dis}} < R_{\mathrm{out}}$ (or at $x^{-}_{\mathrm{dis}}$ where $-R_{\mathrm{out}} < x^{-}_{\mathrm{dis}} < -R^{\prime}_{\mathrm{in}}$), we exclude all $x \geq x^{+}_{\mathrm{dis}}$ (or $x \leq x^{-}_{\mathrm{dis}}$) from the fit. 
In the fit, each velocity measurement is weighted by $N / \sigma^2_{V_{\mathrm{LOS}}}$, where $\sigma_{V_{\mathrm{LOS}}}$ is its uncertainty and $N$ is the number of gas cells in its bin. 
We only fit velocity curves that have at least 4 measurements in both $R^{\prime}_{\mathrm{in}} < |x| < R_{\mathrm{out}}$ regions and at least 1 measurement in the $-R^{\prime}_{\mathrm{in}} < x < R^{\prime}_{\mathrm{in}}$ region. 
One example of a line-of-sight velocity curve and its Equation~\ref{eq:diskfit} fit is displayed in the bottom-right corner of Figure~\ref{fig:subs}. 

We take the maximum line-of-sight velocity standard deviation along the edge-on LOS velocity curve as a measure of the velocity dispersion present in a galaxy. 
We hereafter refer to this measure as $\sigma$ in kinematical contexts. 
While the exact definitions of $\sigma$ vary from study to study (e.g. \citealt{Pillepich19} instead measures $\sigma$ from face-on orientation via the extraplanar motions of gas), it is commonly used in conjunction with $V_c$ by theorists and observers to quantify the degree to which a galaxy is either supported by rotation or dispersion \citep[e.g.][]{Jones13, Leethochawalit16, Ma17, Pillepich19}, as we do in Sections~\ref{sec:gradkine}~and~\ref{sec:dkine}. 

\subsection{Metallicity profiles and gradients}
\label{sec:profgrad}

\begin{figure}
\centerline{\vbox{\hbox{
\includegraphics[width=0.45\textwidth]{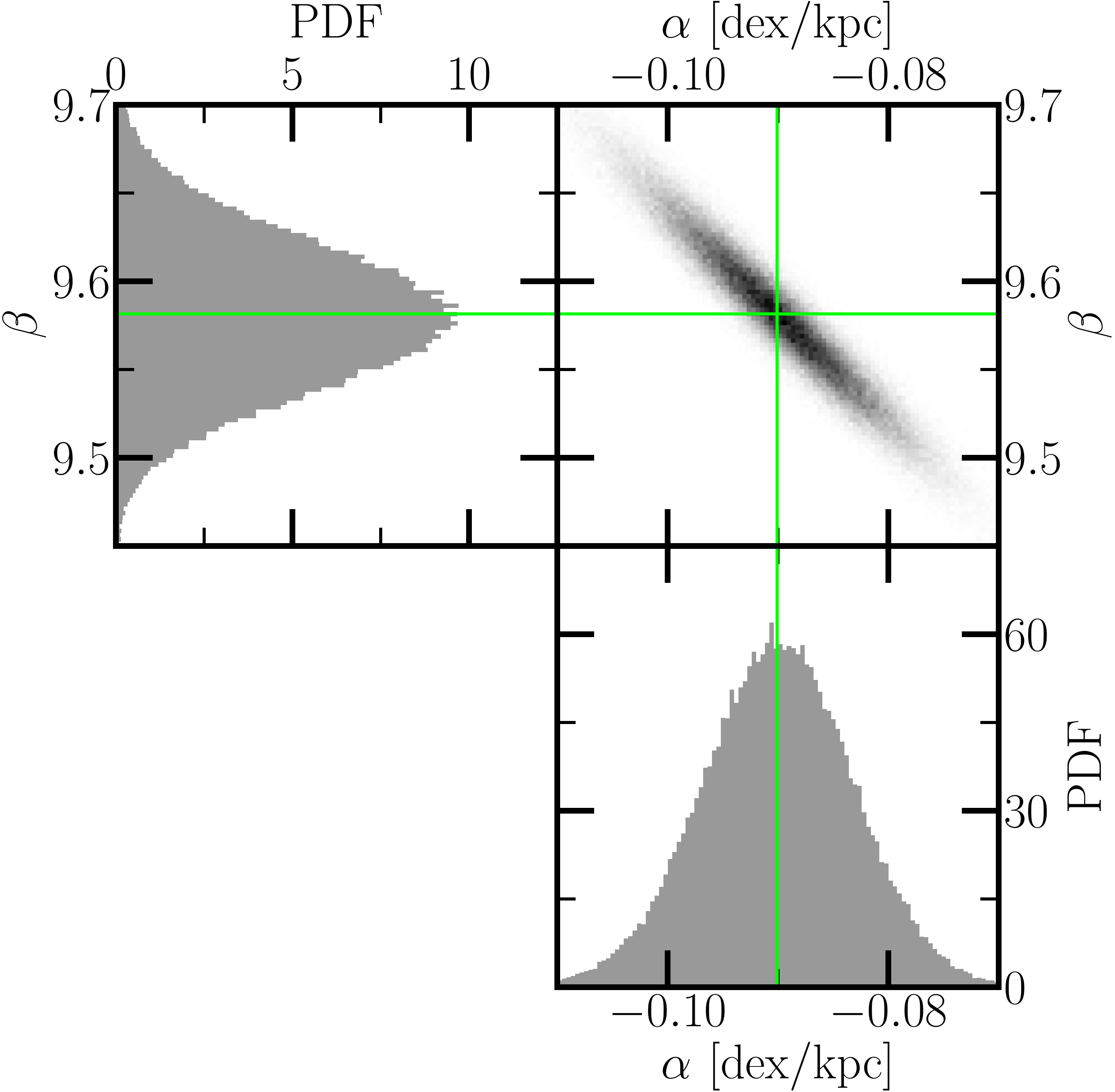} 
}}}
\caption{
The parameter ($\alpha$ and $\beta$) distributions of Equation~(\ref{eq:gradfit}) when fitted to the reduced gas-phase metallicity profile of the galaxy shown in Figure~\ref{fig:subs} via the methods described in Section~\ref{sec:profgrad}. 
$\alpha$ is measured using physical kpc. 
The mean of the $\alpha$ parameter distribution, marked by the vertical green line, is taken as the gas-phase metallicity gradient of the galaxy. 
The mean of the $\beta$ parameter distribution, marked by the horizontal green line, is the extrapolated central metallicity of the galaxy (i.e., the intercept normalization).
}
\label{fig:proffit}
\end{figure}

Our methods of generating galaxy metallicity profiles and measuring metallicity gradients closely follow those of \citetalias{Ma17}. 
We first orient each galaxy such that the gas disk is viewed face-on via the inclination angles described in Section~\ref{sec:galphysprops}. 
This rotation effectively de-projects the radial galactocentric distance of all gas. 
Next, we create maps of galaxy metallicity. 
One example of these metallicity maps is shown in the panel at the top-right corner of Figure~\ref{fig:subs}. 
Each map has a pixel width of $0.1$ kpc and a vertical/horizontal extent of $\pm 2R_{\mathrm{out}}$. 
We exclude any pixels with gas mass surface density $\Sigma_g < 10 \, \Msun \mathrm{pc^{-2}}$ from the following analysis to ensure each pixel is well-populated with gas cells and likely contains detectable \hii\,regions. 
Additionally, to avoid including diffuse gas cells from outside the disk in our analysis, we exclude all gas cells with hydrogen number density $n_{\mathrm{H}} < 0.13\, \mathrm{cm^{-3}}$, the approximate density threshold for star formation in TNG50. 
The value of a given pixel is its oxygen-to-hydrogen abundance ratio, calculated using the total number of oxygen and hydrogen nuclei in all remaining gas cells within that pixel. 
As is customary, we define the abundance ratio $\epsilon$ of species $X$ and hydrogen to be
\begin{equation}
\label{eq:abrat}
    \epsilon \left( N_X, N_{\mathrm{H}} \right) = \log{ \left( N_X / N_{\mathrm{H}} \right) } + 12 
\end{equation}
where $N_X$ and $N_{\mathrm{H}}$ are the number of $X$ and hydrogen nuclei, respectively. 
An example metallicity profile is shown in the bottom-left corner (large panel) of Figure~\ref{fig:subs}. 

We reduce these metallicity profiles to a set of median values sampled with finite resolution. 
At 0.1 kpc intervals of 2D galactocentric distance $r$, we search a $\Delta r = \pm 0.05$ kpc range for at least 16 pixels. 
If at least 16 pixels are not found within the $\Delta r = \pm 0.05$ kpc range, we expand this range until meeting the pixel requirement -- first to $\pm 0.125$ kpc, then $\pm 0.25$ kpc, then $\pm 0.5$ kpc, and finally $\pm 1$ kpc. 
For each $r$, we compute the metallicity median and standard deviation of all pixels within $\Delta r$, along with the standard deviation of their galactocentric distances. 
We take these standard deviations as the uncertainties of each median metallicity value and $r$. 
We hereafter refer to the these profiles as ``reduced metallicity profiles''. 

We find that the reduced metallicity profiles of galactic star-forming regions ($R_{\mathrm{in}} < r < R_{\mathrm{out}}$) are reasonably well-fit by the linear function 
\begin{equation}
\label{eq:gradfit}
    \epsilon \left( r \right) = \alpha r + \beta
\end{equation}
where parameter $\alpha$ is the metallicity gradient and $\beta$ is the extrapolated central metallicity (i.e., the intercept normalization). 
We choose to employ Equation~(\ref{eq:gradfit}) because, in addition to being a generally good fit to our simulated systems (e.g. see the lower left panel of Figure~\ref{fig:subs}), observers typically utilize linear fits to measure metallicity gradients \citep[e.g.][]{Jones10, Yuan11, Swinbank12}. 
Because the \hii\, regions used for metallicity measurements are associated with star formation, we only fit the metallicity profile of the galactic star-forming region (see Section~\ref{sec:galphysprops}). 
Specifically, we fit the region $\left[ R_{\mathrm{in}}^{\prime}, R_{\mathrm{out}} \right]$, where $R_{\mathrm{in}}^{\prime} = R_{\mathrm{in}} + 0.25 \left( R_{\mathrm{out}} - R_{\mathrm{in}} \right)$. 
Following \citetalias{Ma17} (and similar to \citealt{Pilkington12b} and \citetalias{Gibson13}), we do not fit the inner $1/4$ of the galactic star-forming region because of its proximity to the galactic central region, which often possess a gradient that is either much steeper or much flatter than that of the star-forming region. 
We also exclude the metallicity profile at $r > R_{\mathrm{out}}$ from our fit because it often exhibits a flattened gradient -- similar flattening phenomena have been observed in both the stellar and gas-phase metallicity gradients of many local galaxies \citep[e.g.][]{Vlajic09, Vlajic11, Sanchez14}. 
For these reasons, observers also frequently fit a specific region related to some characteristic galactic radius. 
For example, several low-redshift studies \citep[e.g.][]{Sanchez12b, Sanchez14, Sanchez-Menguiano16, Belfiore17} fit the region $\left [0.5R_e, 2R_e \right]$, where $R_e$ is the galactic effective radius. 
However, this practice is more rare for high-redshift studies, as the metallicity profiles measured at these redshifts appear well-approximated by a single linear fit at all galactocentric distances, although this may be a product of limited spatial resolution. 

To obtain the fit parameter distribution of Equation~(\ref{eq:gradfit}) for a given metallicity profile, we utilize bootstrapping techniques based on those of several previous gradient studies \citep[e.g.][]{Kewley10, Rupke10b, Ho15} and repeatedly fit perturbed data sets that are randomly drawn from the reduced metallicity profile. 
We reason that (at least) two factors can combine to introduce uncertainty into the measured gradient measurements: variance in the galactocentric distances sampled by observations, and intrinsic variance in the metallicity of \hii\, regions at some (also uncertain) galactocentric distance. 
Consider the case of a reduced metallicity profile with $N_0$ measurements. 
To include the former source of uncertainty, we randomly draw $N = N_0$ measurements (with replacement) from the reduced metallicity profile. 
To include the latter source of uncertainty, we perturb these $N$ data points (in both $\epsilon$ and $r$) by random Gaussian deviates scaled to the uncertainties of each point. 
We take the means of the resultant fit parameter distributions as the optimal fit parameters, and the standard deviations of the distributions as the fit parameter uncertainties. 
One example of an optimal fit is shown in the metallicity profile (bottom-left corner) of Figure~\ref{fig:subs}. 
The parameter distributions of this fit are displayed in Figure~\ref{fig:proffit}. 

We emphasize that the methods employed in this section do not constitute a mock analysis of TNG50 gradients. 
Several aspects of these methods -- e.g. rotating galaxies to face-on orientation, removing diffuse ionized gas contamination, sampling abundances in 0.1 kpc pixels, and measuring abundances without SFR weighting -- are not realistic given the current limitations of the observing paradigm. 
Nonetheless, our methodology approximates the techniques and challenges of observations in several ways. 
For example, we measure abundances only in pixels containing many gas cells with densities that meet the star formation prescription threshold, and closely follow standard procedures to fit profiles and generate uncertainties. 
Thus, as long as we keep the caveats of our methodology in mind, we believe that it should allow a careful comparison to the current body of gradient measurements. 
We leave a true mock analysis of metallicity profiles and gradients to a future work. 

\section{Results}

\label{sec:results}

\subsection{Local metallicity profiles and gradients}
\label{sec:gradz0}

\begin{figure}
\centerline{
\includegraphics[width=0.45\textwidth]{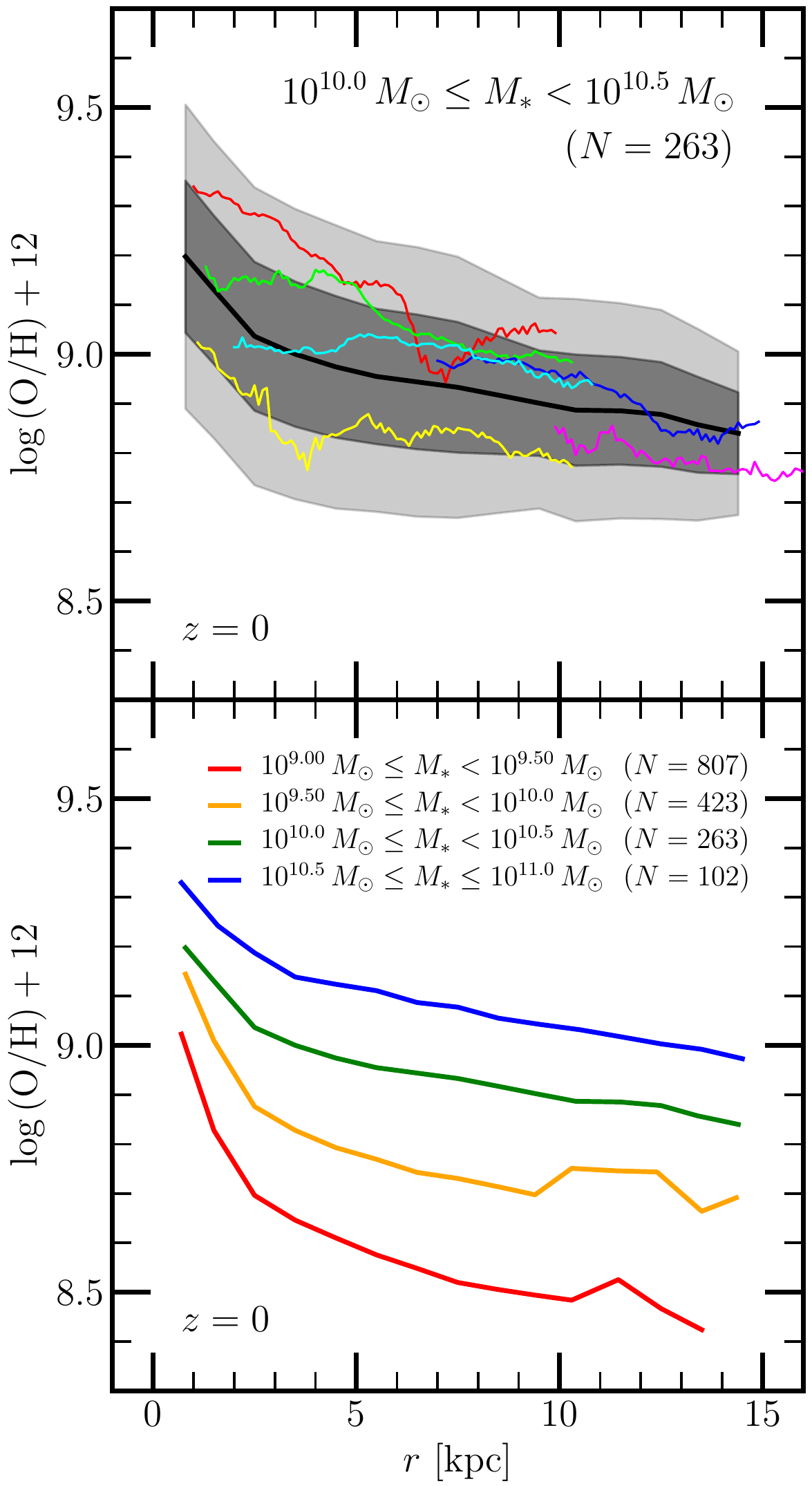}
}
\caption{
\textit{Top:} The median gas-phase metallicity profile generated from the individual reduced gas-phase metallicity profiles in the star-forming regions (see Sections~\ref{sec:galphysprops}~and~\ref{sec:profgrad}) of all redshift $z = 0$ TNG50 star-forming galaxies in the $10^{10} \, M_{\odot} \leq M_* < 10^{10.5} \, M_{\odot}$ mass bin. 
Colored lines show several examples of individual star-forming region reduced metallicity profiles. 
The black line marks the median of all individual profiles, and the shaded regions indicate their $1\sigma$ and $2\sigma$ spread. 
\textit{Bottom:} The median gas-phase metallicity profiles generated from the individual reduced gas-phase metallicity profiles in the star-forming regions (see Sections~\ref{sec:galphysprops}~and~\ref{sec:profgrad}) of redshift $z = 0$ TNG50 star-forming galaxies in all mass bins. 
The median metallicity profiles are separated into four mass bins, as indicated in the legend. 
In both panels, $r$ is measured in physical kpc.
All TNG50 profiles presented in this paper are mass-weighted and measured from face-on orientation.
}
\label{fig:med0}
\end{figure}

Because previous simulations and observations have suggested a possible relation between galaxy stellar mass and metallicity profiles/gradients, we separate the TNG50 sample into four stellar mass bins, ranging from the minimum selected stellar mass ($M_* = 10^9 \, M_{\odot}$) to the maximum ($M_* = 10^{11} \, M_{\odot}$) with an increment of 0.5 dex. 
The top panel of Figure~\ref{fig:med0} shows the median metallicity profile generated from the reduced metallicity profiles in the star-forming regions of all redshift $z = 0$ galaxies in the $10^{10} \, M_{\odot} \leq M_* < 10^{10.5} \, M_{\odot}$ mass bin. 
The median metallicity profile is displayed as a black line, while several examples of reduced metallicity profiles from individual redshift $z = 0$ galaxies are shown as colored lines. 
The shaded regions indicate the $1\sigma$ and $2\sigma$ spread of all individual reduced metallicity profiles in the mass bin. 

We find that the vast majority of gas-phase reduced metallicity profiles decay with radius. 
While there is significant noise among the individual reduced metallicity profiles, Figure~\ref{fig:med0} demonstrates that they remain generally close to the overall median metallicity profile at all $r$, rarely deviating by more than $\sim 0.3$ dex. 
However, we caution that the median metallicity profiles are artificial and should not to be interpreted as the typical individual profiles for each mass bin. 
Because they include only the star-forming region of each galaxy, individual metallicity profiles are approximately linear and seldom span significantly more than $\sim 5$ kpc alone (see e.g. Figure~\ref{fig:subs}). 
Still, sections of the median profiles near some $r$ can be used to predict the approximate slope and normalization of individual profiles at that $r$. 

The bottom panel of Figure~\ref{fig:med0} displays the median metallicity profiles generated from the reduced metallicity profiles in the star-forming regions of all redshift $z = 0$ galaxies separated into the aforementioned four stellar mass bins. 
All mass bins show similar decay in metallicity as a function of radius. 
Although not shown in Figure~\ref{fig:med0}, the spread of the individual metallicity profiles of each mass bin are similar to those of the mass bin shown in Figure~\ref{fig:med0}. 
That is, the reduced metallicity profiles in the star-forming regions of galaxies in all mass bins rarely deviate significantly from the median metallicity profiles shown in Figure~\ref{fig:med0}. 
As required by the MZR, higher mass galaxies have higher metallicity normalization. 
Visual inspection reveals that, at all $r$, the median metallicity gradients are steeper for lower mass galaxies when compared to higher mass galaxies at redshift $z = 0$. 
We will revisit this point later (in Section~\ref{sec:gradm}) when we consider the stellar mass dependence of metallicity gradients and redshift evolution of this dependence. 

\begin{figure}
\centerline{\vbox{\hbox{
\includegraphics[width=0.45\textwidth]{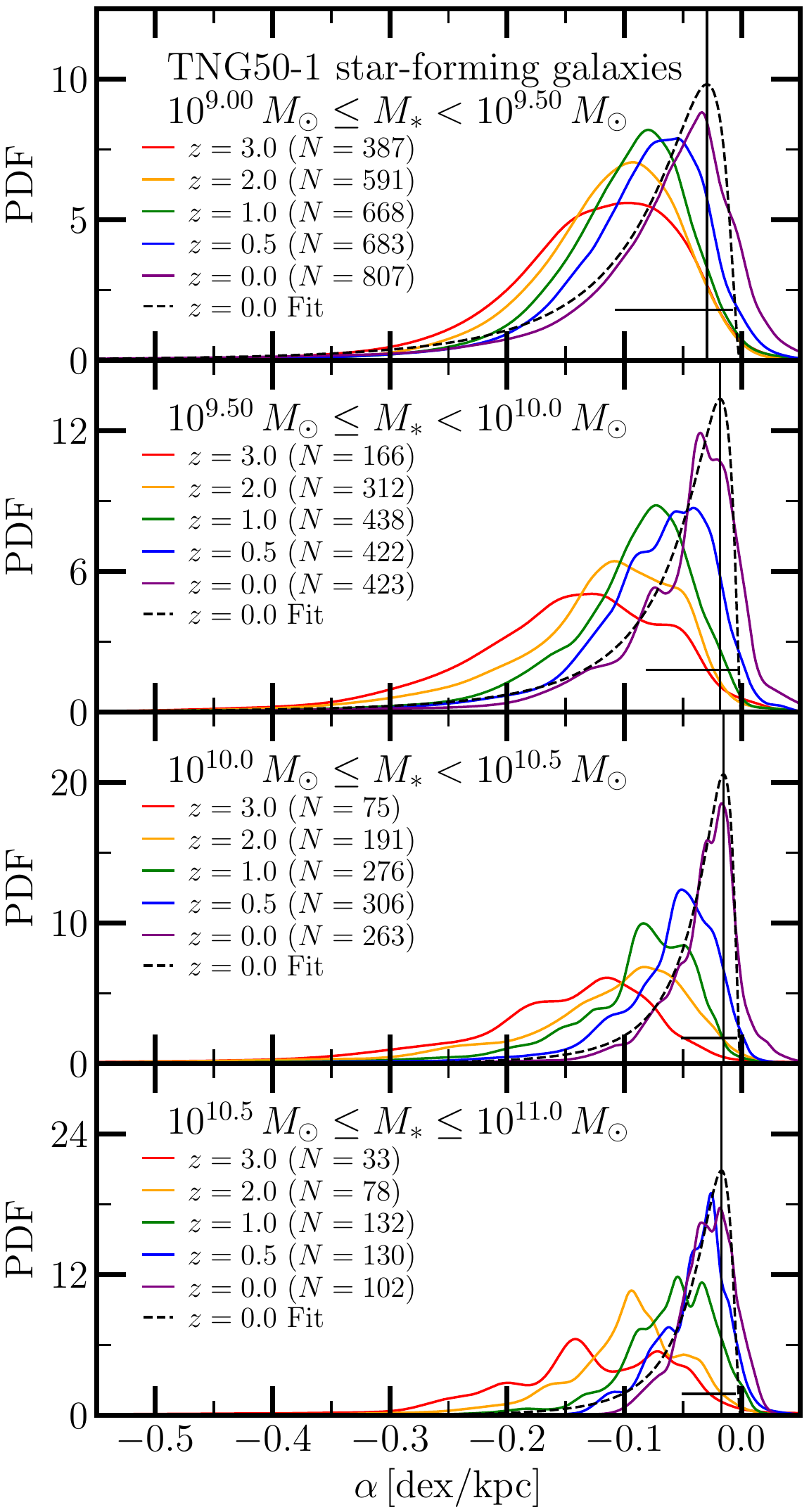}
}}}
\caption{
The metallicity gradient ($\alpha$) distributions of $10^9 \, M_{\odot} \leq M_* \leq 10^{11} \, M_{\odot}$ TNG50-1 star-forming galaxies, separated by stellar mass and redshift. 
Each histogram is fit with a log-normal distribution -- for each mass bin, the redshift $z = 0$ fit is displayed as a black dashed curve. 
The peak of the redshift $z = 0$ log-normal fits are marked by black vertical lines, and the shortest spread around the peak that encloses 68\% of the distribution's probability is given as a horizontal black line. 
The peak and spread of each log-normal fit are used in the following tables and figures to characterize each gradient distribution. 
}
\label{fig:dist}
\end{figure}

Figure~\ref{fig:dist} shows metallicity gradients distributions of all redshift $z = 0$ galaxies separated into the aforementioned four stellar mass bins. 
For all stellar mass bins and redshifts, these distributions are well-approximated by log-normal distributions and we fit them accordingly. 
Because log-normal distributions are only defined on the interval $\left( 0 , \infty \right)$, we exclude all positive gradients from the distribution and fit the negative of the remaining metallicity gradients (i.e., $-\alpha$). 
Very few galaxies in our sample exhibit positive metallicity gradients, so excluding positive gradients does not significantly alter the gradient distributions. 
Hereafter, we characterize these distributions by their peak and the shortest spread around this peak that encloses 68\% of the distribution. 

\subsection{Metallicity gradient redshift evolution}
\label{sec:gradz}

\begin{center}
\begin{table}                                                                    
    \centering                                                                   
    \caption{The redshift evolution of metallicity gradients in the selected TNG50 star-forming galaxy sample (see Section~\ref{sec:galselection}), separated into four stellar mass bins.
    The quoted metallicity gradients and their uncertainties are the peak and spread of log-normal fits to the gradient distributions of each mass bin at each redshift (see Section~\ref{sec:gradz0}~and~Figure~\ref{fig:dist}).}
    \label{tab:gradvz}                                                           
    \begin{tabular}{rrrrr}                                                       
        \hline                                                                   
        $z$ & $M_*^{\mathrm{min}}$ & $M_*^{\mathrm{max}}$ & $N$ & $\alpha$ \\ 
            & $\left[ \log \left( \frac{M_*}{M_{\odot}} \right) \right]$ & $\left[ \log \left( \frac{M_*}{M_{\odot}} \right) \right]$ & & $\left[ \mathrm{dex / kpc} \right]$ \\ 
        \hline                                                                   
            $3.0$ & $ 9.0$ & $ 9.5$ & $387$ & $-0.084^{+0.045}_{-0.097}$ \\ [2pt] 
            $3.0$ & $ 9.5$ & $10.0$ & $166$ & $-0.102^{+0.054}_{-0.114}$ \\ [2pt] 
            $3.0$ & $10.0$ & $10.5$ & $ 75$ & $-0.111^{+0.051}_{-0.094}$ \\ [2pt] 
            $3.0$ & $10.5$ & $11.0$ & $ 33$ & $-0.085^{+0.047}_{-0.108}$ \\ [2pt] 
            $2.0$ & $ 9.0$ & $ 9.5$ & $591$ & $-0.082^{+0.040}_{-0.077}$ \\ [2pt] 
            $2.0$ & $ 9.5$ & $10.0$ & $312$ & $-0.079^{+0.042}_{-0.090}$ \\ [2pt] 
            $2.0$ & $10.0$ & $10.5$ & $191$ & $-0.070^{+0.039}_{-0.090}$ \\ [2pt] 
            $2.0$ & $10.5$ & $11.0$ & $ 78$ & $-0.076^{+0.034}_{-0.063}$ \\ [2pt] 
            $1.0$ & $ 9.0$ & $ 9.5$ & $668$ & $-0.072^{+0.035}_{-0.067}$ \\ [2pt] 
            $1.0$ & $ 9.5$ & $10.0$ & $438$ & $-0.064^{+0.033}_{-0.066}$ \\ [2pt] 
            $1.0$ & $10.0$ & $10.5$ & $276$ & $-0.060^{+0.030}_{-0.060}$ \\ [2pt] 
            $1.0$ & $10.5$ & $11.0$ & $132$ & $-0.038^{+0.022}_{-0.051}$ \\ [2pt] 
            $0.5$ & $ 9.0$ & $ 9.5$ & $683$ & $-0.053^{+0.031}_{-0.072}$ \\ [2pt] 
            $0.5$ & $ 9.5$ & $10.0$ & $422$ & $-0.040^{+0.026}_{-0.069}$ \\ [2pt] 
            $0.5$ & $10.0$ & $10.5$ & $306$ & $-0.035^{+0.021}_{-0.052}$ \\ [2pt] 
            $0.5$ & $10.5$ & $11.0$ & $130$ & $-0.024^{+0.015}_{-0.036}$ \\ [2pt] 
            $0.0$ & $ 9.0$ & $ 9.5$ & $807$ & $-0.029^{+0.021}_{-0.079}$ \\ [2pt] 
            $0.0$ & $ 9.5$ & $10.0$ & $423$ & $-0.018^{+0.014}_{-0.063}$ \\ [2pt] 
            $0.0$ & $10.0$ & $10.5$ & $263$ & $-0.015^{+0.011}_{-0.036}$ \\ [2pt] 
            $0.0$ & $10.5$ & $11.0$ & $102$ & $-0.017^{+0.011}_{-0.033}$ \\ [2pt] 
        \hline                                                                   
    \end{tabular}                                                                
\end{table}                                                                      
\end{center}

\begin{figure*}
\centerline{\vbox{
\includegraphics[width=\textwidth]{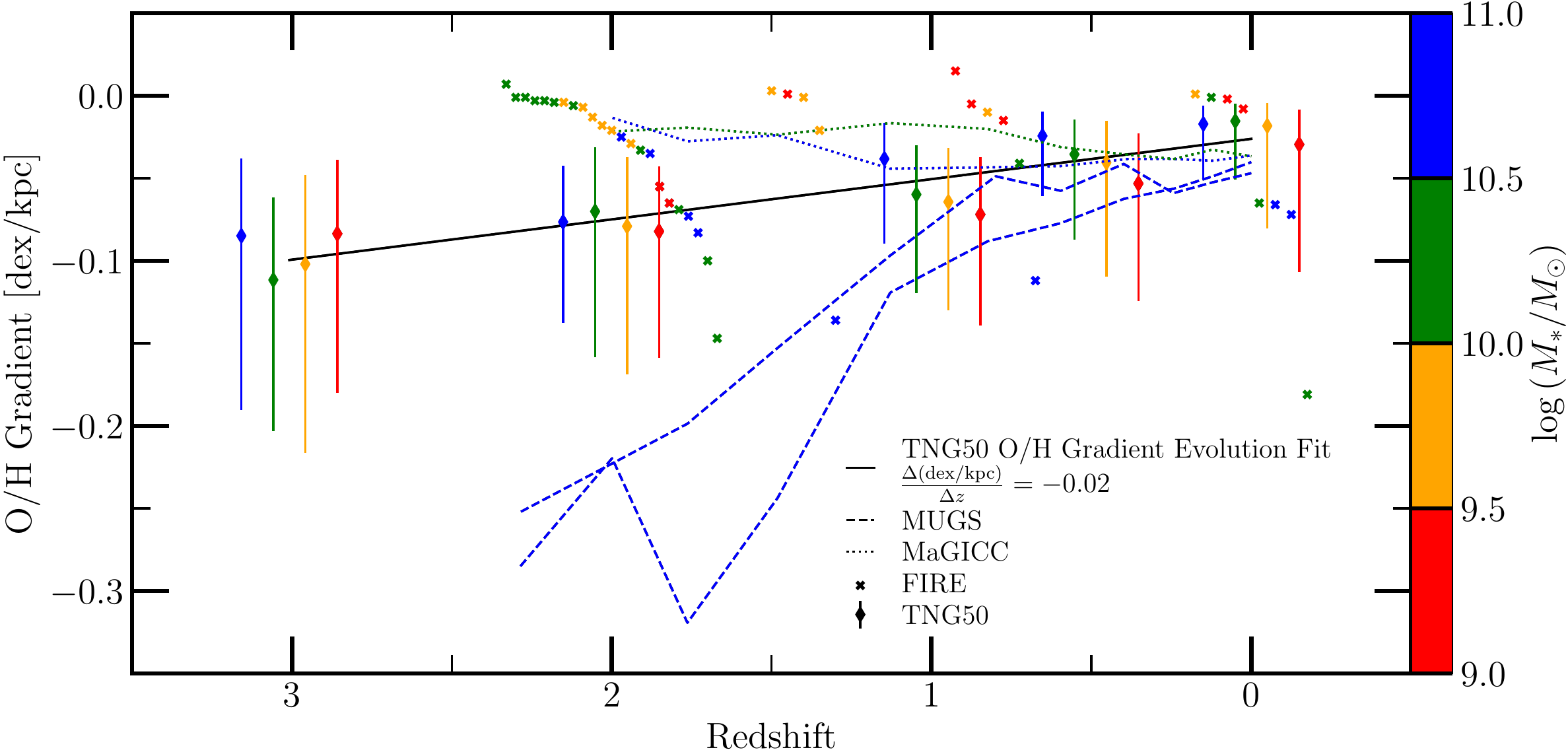}
}}
\caption{
Gas-phase metallicity gradients as a function of redshift for $10^{9}\,M_{\odot} \leq M_* \leq 10^{11}\,M_{\odot}$ TNG50 star-forming galaxies at redshifts $z=$0--3. 
Each data point shows peak and spread of a log-normal fit to the metallicity gradient distribution of a given mass bin (see Section~\ref{sec:gradz0}, Figure~\ref{fig:dist}, and Table~\ref{tab:gradvz}). 
The black solid line indicates the least-squares fit to the redshift evolution of all mass bins, which has a slope of $-0.02 \, \left[ \mathrm{dex \, kpc^{-1}} / \Delta z \right]$. 
We compare the redshift evolution of TNG50 gradients to that of MUGS and MaGICC simulated galaxies (systems g1536 and g15784 from the MUGS simulation suite; \citealt{Gibson13}), as well as FIRE simulated galaxies \citep{Ma17}.
Each galaxy/bin is color-coded by stellar mass. 
TNG50 and FIRE data points are artificially offset in redshift to avoid stacking. 
The true redshifts of the TNG50 data are $z = 3, 2, 1, 0.5, 0$, and the FIRE data $z = 2, 1.4, 0.8, 0$. 
}
\label{fig:gibssim}
\end{figure*}

\begin{figure*}
\centerline{\vbox{
\includegraphics[width=1.0\textwidth]{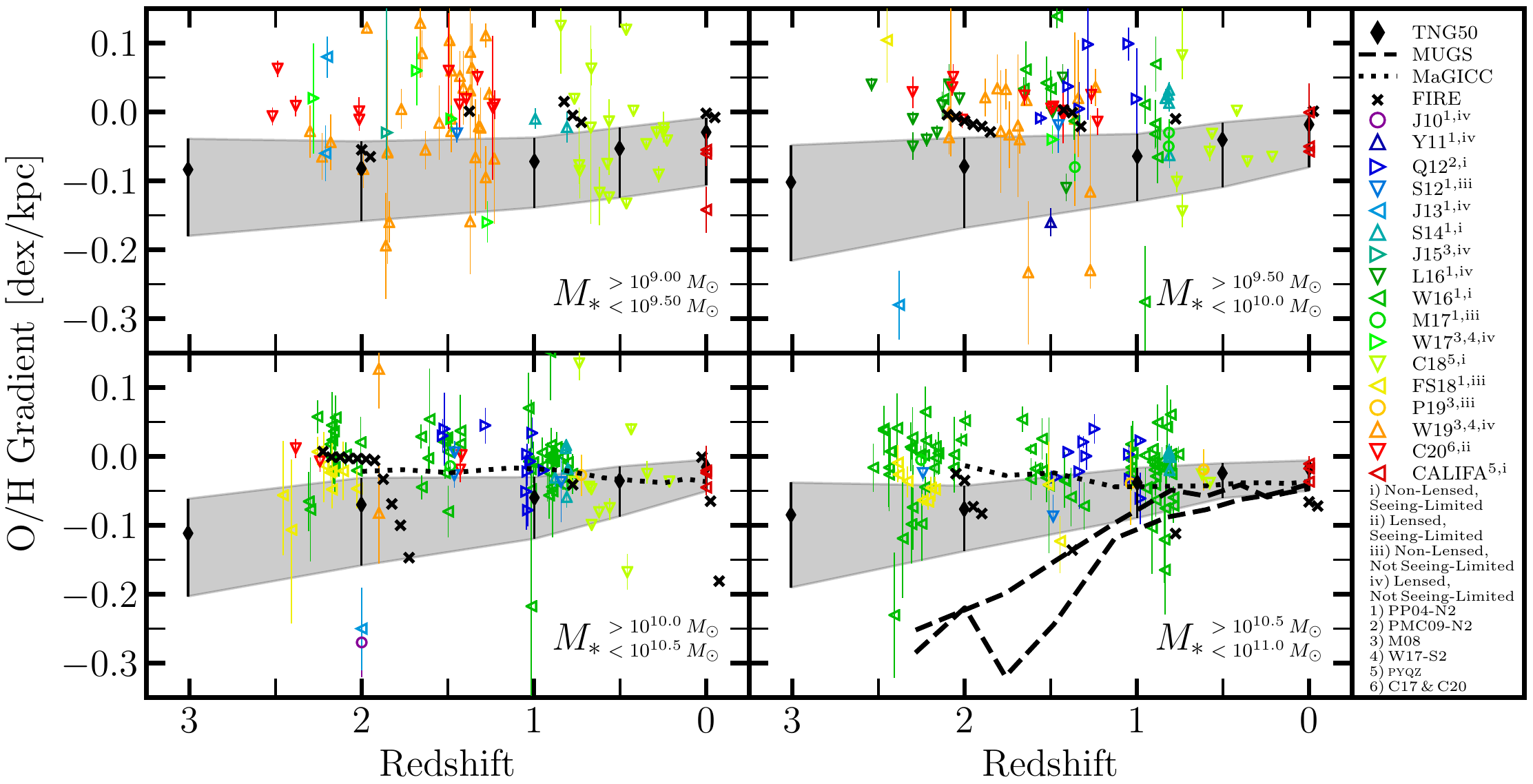}
}}
\caption{
The gas-phase metallicity gradient redshift evolution of $10^{9}\,M_{\odot} \leq M_* \leq 10^{11}\,M_{\odot}$ TNG50 star-forming galaxies from redshifts $z=$0--3 compared at face value to observations of gradients in galaxies within the same stellar mass and redshift range.
Each panel displays the gradient evolution of a different stellar mass bin, as indicated in each bottom-right corner. 
Black diamonds and shaded regions show the peak and spread of the gradient distributions of each TNG50 mass bin. 
The colored data points indicate observations from \citet{Jones10}, \citet{Yuan11}, \citet{Queyrel12}, \citet{Swinbank12}, \citet{Jones13}, \citet{Stott14}, \citet{Jones15}, \citet{Leethochawalit16}, \citet{Wuyts16}, \citet{Molina17}, and \citet{Wang17}, \citet{Carton18}, \citet{ForsterSchreiber18}, \citet{Patricio19}, \citet{Wang19}, \citet{Curti20b}, and CALIFA \citep[][]{Sanchez12b, Sanchez14, Sanchez-Menguiano16} as indicated in the legend. 
Also in the legend, we state whether each galaxy sample was lensed or non-lensed and whether observations were or were not seeing-limited, as well as the indicator/calibrator used by each study to produce the data shown here. 
These indicators/calibrators used are PP04-N2 \citep[][]{PP04}, PMC09-N2 \citep[][]{Perez-Montero09}, M08 \citep[][]{Maiolino08}, W17-S2 \citep[][]{Wang17}, \textsc{pyqz} \citep[][]{Dopita13}, C17 \citep[][]{Curti17}, and C20 \citep[][]{Curti20a}. 
We note that some studies \citep[e.g.][]{Wuyts16, ForsterSchreiber18} only share the measured gradient of N2 -- to infer a metallicity gradient for these data, we use the PP04-N2 calibrator. 
We also show gradients from the MUGS and MaGICC simulations of \citet{Gibson13} and from the FIRE simulations of \citet{Ma17}. 
FIRE data points are artificially offset in redshift to avoid stacking -- the true redshifts of these data are $z = 2, 1.4, 0.8, 0$. 
Simulated gradients are measured from the metallicity profiles of galaxies projected face-on.
}
\label{fig:gibsobs}
\end{figure*}

Figure~\ref{fig:gibssim} shows the redshift evolution of the metallicity gradient distributions for all four distinct mass bins from redshifts $z = $ 0 -- 3. 
Each data point gives the peak and spread of a log-normal fit to the metallicity gradient distribution for each mass bin at each redshift (see Figure~\ref{fig:dist}). 
The exact evolution of each mass bin is quoted in Table~\ref{tab:gradvz}. 

In general, we observe a monotonic decrease in metallicity gradient steepness with decreasing redshift. 
This decrease in gradient steepness exists for each individual mass bin and for the population as a whole. 
Further, we find that the redshift evolution of gradients is very similar across mass bins. 
We use linear regression to fit a single line through the gradient evolution of the entire population (shown by the black line in Figure~\ref{fig:gibssim}), which gives a metallicity gradient evolution rate of $-0.02 \left[ \mathrm{dex \, kpc^{-1}} / \Delta z \right]$. 
The gradient evolution rates of individual mass bins do not significatly deviate from this overall rate. 
As we will discuss in Section~\ref{sec:gradr}, gradient steepness increasing with redshift seems related to galaxy disk sizes decreasing with redshift. 

Inspection of the error bars in Figure~\ref{fig:gibssim} also reveals that the spread of gradient distributions increases with redshift. 
That is, we observe a larger diversity of gradients at high redshift, with distributions spanning from approximately $\alpha = -0.2$ dex/kpc to $\alpha = -0.05$ dex/kpc at redshift $z = 3$. 
However, we note that the gradients of all mass bins exhibit significant diversity at all redshifts. 
Even at redshift $z = 0$, where the gradient distributions are tightest, the distribution spread stretches from roughly $\alpha = -0.1$ dex/kpc to $\alpha = -0.01$ dex/kpc. 
This positive correlation between gradient distribution spread and redshift holds true for the galaxy population as a whole and for each individual stellar mass bin. 
We find that the most massive bin considered in our analysis ($10^{10.5} \, M_{\odot} \leq M_* \leq 10^{11} \, M_{\odot}$, shown as blue in Figure~\ref{fig:gibssim}) exhibits the strongest redshift evolution in gradient distribution spread, although this mass bin is somewhat underpopulated at high redshift. 
The next most massive bin ($10^{10} \, M_{\odot} \leq M_* < 10^{10.5} \, M_{\odot}$, shown as green in Figure~\ref{fig:gibssim}) is significantly more populated at high redshift and also exhibits very strong redshift evolution in gradient distribution spread. 
All mass bins begin with gradient distribution spreads of $\sim 0.2$ dex/kpc at redshift $z = 3$. 
The gradient distribution spread of the second-most massive bin shrinks to $\sim 0.05$ dex/kpc by redshift $z = 0$, while the least massive bin still has a distribution spread of $\sim 0.1$ dex/kpc. 

To compare our results with current observations, Figure~\ref{fig:gibsobs} shows the metallicity gradient evolution of our TNG50 sample along with observed gradients from a collection of relevant studies -- \citet{Jones10}, \citet{Yuan11}, \citet{Queyrel12}, \citet{Swinbank12}, \citet{Jones13}, \citet{Stott14}, \citet{Jones15}, \citet{Leethochawalit16}, \citet{Wuyts16}, \citet{Molina17}, \citet{Wang17}, \citet{Carton18}, \citet{ForsterSchreiber18}, \citet{Patricio19}, \citet{Wang19}, \citet{Curti20b}, and CALIFA \citep[][]{Sanchez12b, Sanchez14, Sanchez-Menguiano16}. 
Additionally, we include the results of several simulations, including FIRE \citepalias{Ma17} and MUGS/MaGICC \citepalias[][]{Gibson13}.
These gradients are separated into four separate panels based on galaxy stellar masses. 

Based on our analysis, TNG50 predicts redshift $z = 0$ metallicity gradients spanning from approximately $-0.1$ dex/kpc to $-0.01$ dex/kpc with a median of roughly $-0.02$ dex/kpc. 
This prediction is consistent with the metallicity gradients measured in local galaxies by CALIFA \citep[][]{Sanchez12b, Sanchez14, Sanchez-Menguiano16}. 
Additionally, we note that TNG50 is not unique in this prediction. 
The MUGS, MaGICC, and FIRE simulations all predict similar metallicity gradients at redshift $z = 0$, as shown in Figure~\ref{fig:gibsobs}. 
However, there is one striking difference between the low-redshift gradients of TNG and FIRE. 
Apparent in both Figures~\ref{fig:gibssim}~and~\ref{fig:gibsobs}, TNG generally predicts that lower-mass galaxies have steeper gradients, while FIRE predicts that higher-mass galaxies have steeper gradients. 
This disagreement is likely a result of fundamental differences between the TNG and FIRE feedback models, which we discuss in Section~\ref{sec:dsim}. 

In contrast to the agreement among and between simulations and observations regarding metallicity gradients at redshift $z = 0$, there exists some tension among and between simulations and observations at higher redshifts. 
The observational picture of high-redshift gradients has evolved significantly over the past decade. 
While many initial small-sample measurements of high redshift gradients suggested a steepening of gradients at high redshift~\citep[][]{Jones10, Yuan11, Jones13}, \citet{Swinbank12} found high-redshift gradients to be either consistent with or flatter than local gradients, and \citet{Stark08} and \citet{Cresci10} measured inverted high-redshift gradients. 
However, recent years have brought more agreement among high-redshift gradient observations.
Follow-up observations of more sizable populations \citep[e.g.][]{Leethochawalit16, Wuyts16, Wang17, ForsterSchreiber18, Wang19, Curti20b} tend to agree with \citet{Stark08}, \citet{Cresci10}, and \citet{Swinbank12}, finding no clear evidence for a significant increase in gradient steepness. 
As noted previously, TNG50 predicts a steady increase in metallicity gradient steepness with redshift at a rate of $\sim 0.02 \left[ \mathrm{dex \, kpc^{-1}} / \Delta z \right]$ (as shown by the black line in the left panel of Figure~\ref{fig:gibssim}). 
This TNG50 prediction is broadly consistent with those observations that indicate steeper gradients at higher redshift, but fails to capture the abundance of flat and inverted gradients for $z \gtrsim 1$. 
The high-redshift gradients predicted by TNG50 are generally more steep than those of FIRE and MaGICC and more shallow than those of MUGS.
We further compare the metallicity gradients measured by simulations and observations in Sections~\ref{sec:dsim}~and~\ref{sec:dobs}. 

\subsection{Galaxy stellar mass vs. metallicity gradient}
\label{sec:gradm}
\begin{figure*}
\centerline{\vbox{
\includegraphics[width=\textwidth]{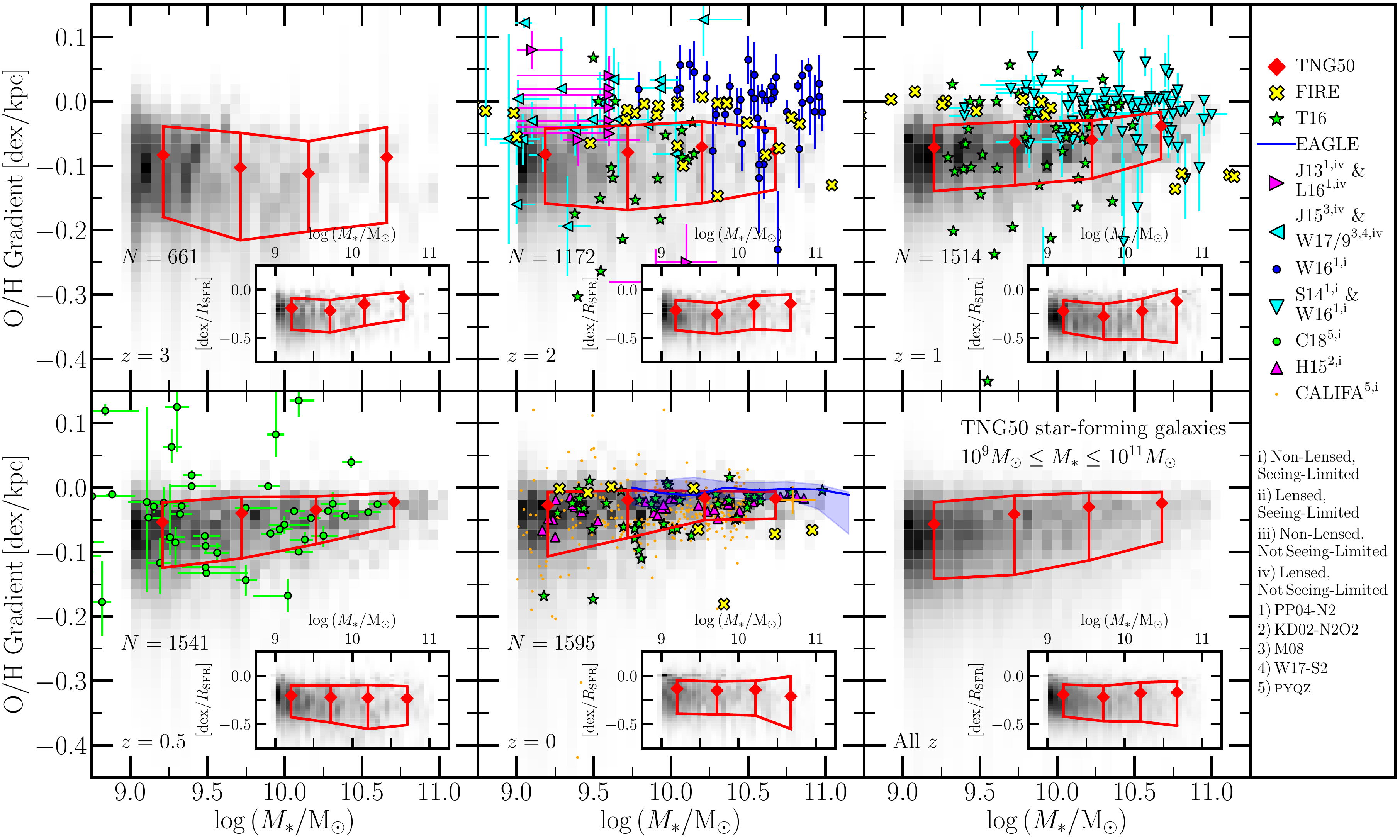}
}}
\caption{
The evolution through cosmic time of the relation between metallicity gradient ($\alpha$, see Section~\ref{sec:profgrad}) and galaxy stellar mass ($M_*$). 
Each panel displays the gradient--$M_*$ 2D histogram of all galaxies in the selected TNG50 sample (see Section~\ref{sec:galselection}) at the redshift noted in each lower-left corner. 
Inset panels show the gradient--$M_*$ 2D histogram for gradients normalized by \RSFR. 
Red points and vertical red lines show the peak and spread of the four log-normal distributions fit to the metallicity gradient distributions of the four selected stellar mass bins (see Section~\ref{sec:gradz0}~and~Figure~\ref{fig:dist}). 
Horizontal red lines, drawn to connect the spread of the gradient distribution fits, display the interpolated envelope of the gradient--$M_*$ relation at each redshift. 
Each panel also shows gradient--$M_*$ data from observations and other simulations, as indicated in the legend. 
The observations included at each redshift are: $z \sim 0$ \citep{Sanchez12b, Sanchez14, Ho15, Sanchez-Menguiano16}, $z \sim 0.5$ \citep{Carton18}, $z \sim 1$ \citep{Stott14, Wuyts16}, and $z \sim 2$ \citep{Jones13, Jones15, Leethochawalit16, Wuyts16, Wang17, Wang19}. 
The legend indicates the spatial resolution of and indicator(s)/calibrator(s) used by each study. 
Observations that (i) use the same indicator(s)/calibrator(s) and (ii) have similar spatial resolution are grouped together \citep[e.g.][]{Stott14, Wuyts16}. 
We include simulation data from \citet{Tissera16}, FIRE \citep{Ma17}, and EAGLE \citep{Tissera19}. 
}
\label{fig:mstar}
\end{figure*}

Following the methodology described in Section~\ref{sec:gradz0} and shown in Figure~\ref{fig:dist}, we fit the metallicity gradient distributions of all stellar mass bins with log-normal distributions at each redshift ($z = $ 3, 2, 1, 0.5, and 0) and present our results in Figure~\ref{fig:mstar}. 

At all redshifts, we indeed find correlations between galaxy stellar mass and metallicity gradient. 
At the highest redshift ($z = 3$), we observe a weak positive correlation between the two parameters. 
It is important to note that, at redshifts $z \gtrsim 2$, the bin containing the most massive galaxies ($10^{10.5} \, M_{\odot} < M_* < 10^{11} \, M_{\odot}$) is sparsely populated (with $N = 33$ and $N = 78$ at redshifts $z = 3$ and $z = 2$, respectively) because such massive galaxies have not had sufficient time to develop. 
Therefore, the data points at high redshift for this mass bin in Figure~\ref{fig:mstar} and in all subsequent figures are somewhat uncertain, and we give less weight to these data points while drawing our conclusions for correlations at early times. 
Keeping this in mind, at redshift $z = 2$, we note that the weak positive correlation between galaxy stellar mass and gradient steepness reverses, becoming a weak negative correlation. 
At redshift $z = 1$, this correlation remains negative and becomes stronger. 
Also at this redshift, we note the beginning of a clear negative correlation between the spread of metallicity gradient distributions and galaxy stellar mass. 
That is, galaxies of lower stellar mass exhibit much more varied metallicity gradients than galaxies of higher stellar mass. 
We also note that, by redshift $z = 1$, the highest-mass bin has become significantly more populated and must now be fully considered in our analysis. 
At redshift $z = 0.5$, the correlations between galaxy stellar mass and both metallicity gradient steepness and diversity are at their strongest. 
At redshift $z = 0$, both correlations remain negative but begin to weaken. 
This is especially true for galaxies of higher stellar mass, for which gradients start to become predominately flat at these late times (likely due to AGN feedback and increasing galaxy size, see Section~\ref{sec:gradr}). 

We also combine the metallicity gradient distributions across all redshifts, creating an ``all-$z$'' gradient distribution for each mass bin. 
We fit these all-$z$ gradient distributions with log-normal distributions and show our results in the bottom-right panel of Figure~\ref{fig:mstar}. 
The aforementioned negative correlations between galaxy stellar mass and both gradient steepness and diversity remain apparent among these all-$z$ distributions. 

The gradients of galaxies in the lower mass bins at redshifts $z = 2$ and $z = 3$ are significantly weaker than the all-$z$ negative gradient--$M_*$ correlation would predict. 
As previously mentioned, there is even an apparent positive gradient--$M_*$ correlation at $z = 3$. 
The comparatively weak gradients of lower-mass galaxies at early times are likely caused by the significant velocity dispersion -- and consequently low $V_c / \sigma$ parameters -- in these young and rapidly star-forming systems. 
This velocity dispersion serves to radially mix the ISM of these systems, flattening their metallicity gradients. 
We further discuss the relation between metallicity gradients and galaxy kinematics in Sections~\ref{sec:gradkine}~and~\ref{sec:dkine}. 

In the inset panels of Figure~\ref{fig:mstar}, we compare galaxy stellar mass and metallicity gradients normalized by galaxy size. 
Specifically, we normalize gradients by \RSFR, the galactic 3D radius that encloses 50\% of star formation. 
We find \RSFR\,to be significantly correlated with metallicity gradient (as described in Section~\ref{sec:gradr}), and suspect that the comparatively weaker negative correlations between gradient and stellar mass may simply be a product of a positive correlation between stellar mass and galaxy size. 
Indeed, the inset panels of Figure~\ref{fig:mstar} show that the weak correlations between physical-scale gradients and stellar mass do not exist for normalized gradients, suggesting that there is not an intrinsic correlation between stellar mass and gradient. 
Similar results are found in several studies of nearby galaxies \citep[e.g.][]{Sanchez12b, Sanchez14, Ho15, Sanchez-Menguiano16}. 

Figure~\ref{fig:mstar} also includes data from observations \citep[][]{Sanchez12b, Sanchez14, Jones13, Stott14, Ho15, Jones15, Leethochawalit16, Sanchez-Menguiano16, Wuyts16, Wang17, Carton18, Wang19}, along with simulation data \citet{Tissera16}, FIRE \citepalias[][]{Ma17}, and EAGLE \citep[][]{Tissera19}. 
We examine the agreements and disagreements between these data and the predictions of TNG50 in Section~\ref{sec:dmobs}. 

\subsection{Galaxy size vs. metallicity gradient}
\label{sec:gradr}

\begin{figure*}
\centerline{\vbox{
\includegraphics[width=\textwidth]{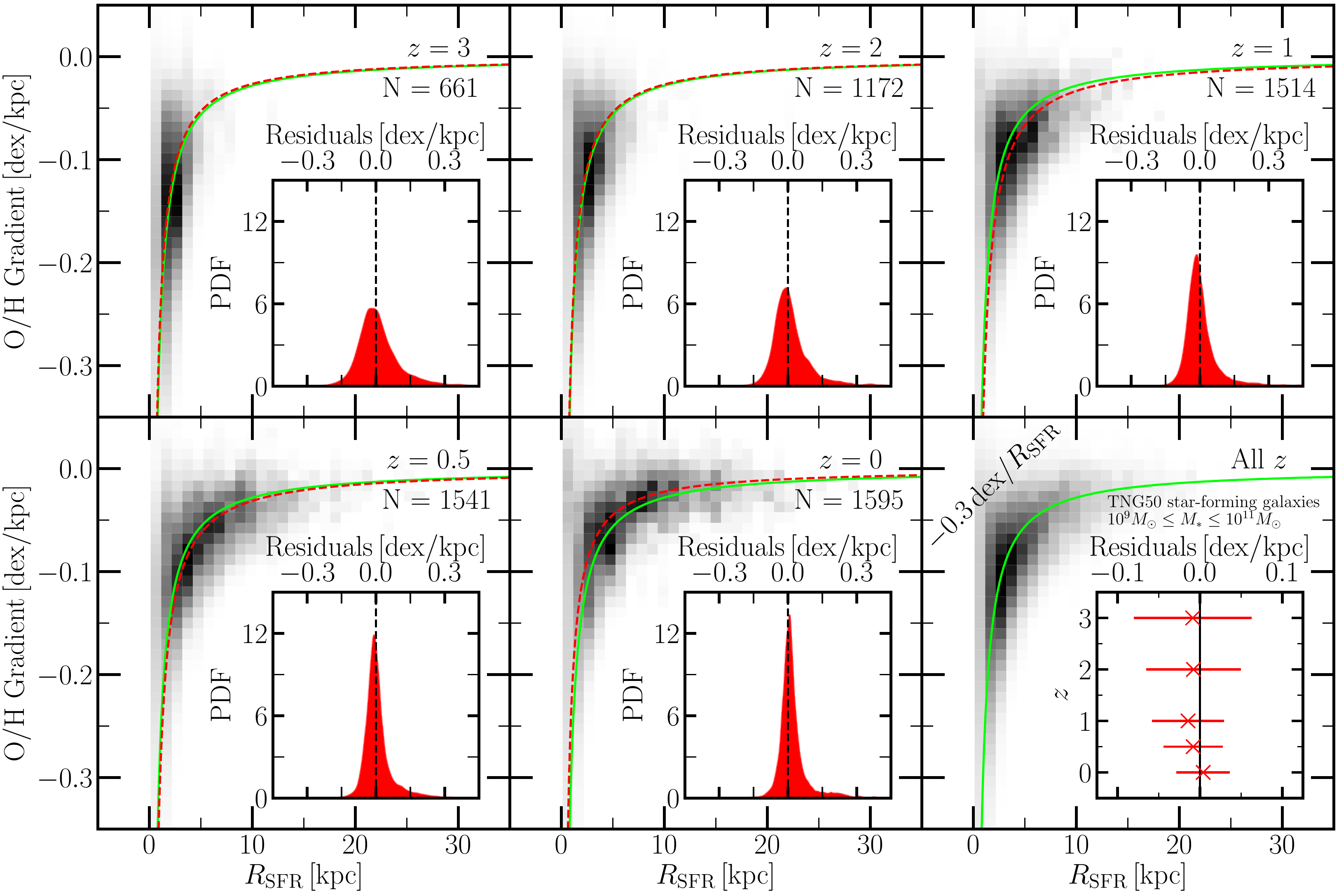}
}}
\caption{
The evolution through cosmic time of the correlation between gas-phase metallicity gradient ($\alpha$) and the 50\% integrated SFR galactic radius ($R_{\mathrm{SFR}}$, see Sections~\ref{sec:galphysprops}~and~\ref{sec:profgrad}). 
Each panel displays the gradient--\RSFR\, 2D histogram of all rotationally-supported galaxies (i.e., $V_c / \sigma > 1$, see Section~\ref{sec:galkinematics}) in the selected TNG50 sample (see Section~\ref{sec:galselection}) at the redshift noted in each upper-right corner. 
Dashed red lines show the least-squares fit of Equation~\ref{eq:corr} to the galaxies at each individual redshift, while solid green lines show this fit to galaxies at all redshifts (i.e., the fit to the bottom-right panel, hereafter the all-$z$ fit). 
Inset panels show the residuals of galaxy metallicity gradients to the all-$z$ fit at each individual redshift, which are best approximated by Gaussian distributions and are fit accordingly. 
The inset plot of the final panel gives the mean and standard deviation of these Gaussian fits to the residuals. 
In all inset panels, the all-$z$ fit is marked by a black line.
}
\label{fig:corr}
\end{figure*}

After analyzing many galactic physical parameters (half-mass stellar/gas radius, gas mass, gas fraction, SFR/sSFR, half-mass stellar age) we find the parameter that most strongly correlates with metallicity gradient is $R_{\mathrm{SFR}}$. 
As defined in Section~\ref{sec:galphysprops}, \RSFR\,is the 3D distance from the center of the galaxy for which half the total SFR of a galaxy is enclosed. 
Figure~\ref{fig:corr} shows a comparison of galaxy metallicity gradients and \RSFR\,for galaxies at redshifts $z = $ 3, 2, 1, 0.5, and 0. 
Each panel displays a 2D histogram of metallicity gradients and \RSFR\,at the redshift noted in each upper-right corner. 

The bottom-right panel of Figure~\ref{fig:corr} displays the gradient--\RSFR\,data for all rotationally-supported galaxies at all redshifts. 
We hereafter refer to this panel as the ``all-$z$'' panel, and to its data as the ``all-$z$'' data. 
The all-$z$ data shows a significant correlation between metallicity gradient and \RSFR. 
We find that this correlation can be fit by the function 
\begin{equation}
\label{eq:corr}
    \alpha \left(R_{\mathrm{SFR}} \right) = -\frac{C}{R_{\mathrm{SFR}}}
\end{equation}
where $\alpha$ is the metallicity gradient and C, the only free parameter, is some constant. 
We fit Equation~\ref{eq:corr} to the all-$z$ data using orthogonal distance regression and obtain a proportionality constant $C = 0.28$\footnote{For $\alpha$ measured in units of $\left[ \mathrm{dex/kpc} \right]$ and \RSFR\, in $\left[ \mathrm{kpc} \right]$, $C$ is dimensionless.}, meaning that metallicity profiles in TNG50 galactic star-forming regions have a characteristic normalized gradient of $\sim -0.3$ dex/\RSFR. 
This fit (the ``all-$z$'' fit) is shown as a solid green line in all panels of Figure~\ref{fig:corr}. 

The other panels of Figure~\ref{fig:corr} display the gradient--\RSFR\,data from galaxies at each individual redshift as noted in the upper-right corner of each panel. 
It should be immediately apparent to the reader that the gradient--\RSFR\,correlation is preserved at each individual redshift. 
Accordingly, we fit Equation~\ref{eq:corr} to the gradient--\RSFR\, data at each individual redshift, again using orthogonal distance regression, and show the resultant fits as dashed red lines in each individual-redshift panel of Figure~\ref{fig:corr}. 
By comparing the fits of Equation~\ref{eq:corr} for each individual redshift (dashed red lines) to the all-$z$ fit (solid green lines) in Figure~\ref{fig:corr}, one will notice that the two fits do not differ significantly at any redshift. 
At most, the two fits deviate by $\Delta C / \sigma = 0.5$ (occuring at redshift $z = 1$). 
Thus, $C$ is not dependent on redshift, meaning that the gradient--\RSFR\, correlation is invariant with time from redshift $z =$ 3 to 0. 
This time-invariance is emphasized by the inset plots of Figure~\ref{fig:corr}, which show the residuals of the all-$z$ fit deviating minimally from zero. 

We note that Figure~\ref{fig:corr} includes only rotationally-supported galaxies (i.e., those with $V_c / \sigma$ > 1, see Section~\ref{sec:galkinematics}~and~Section~\ref{sec:gradkine}) -- to be clear, this cut is performed only in Figure~\ref{fig:corr}. 
We exclude galaxies dominated by dispersion because they often exhibit irregular morphologies and very concentrated regions of star formation. 
Their sizes are therefore not accurately characterized by \RSFR. 
However, as Figure~\ref{fig:kine} shows, $z = 3$ and $z = 2$ are the redshifts at which our sample contains significant populations of galaxies with $V_c / \sigma < 1$. 
Still, even at redshift $z = 3$, most galaxies are rotationally-supported. 
Thus, nearly all of our galaxy sample is included in Figure~\ref{fig:corr}, although we warn that conclusions drawn from this figure do not apply to dispersion-dominated galaxies. 

We further discuss the gradient--\RSFR\, correlation in Section~\ref{sec:dgradrobs}. 

\subsection{Galaxy kinematics vs. metallicity gradient}
\label{sec:gradkine}

\begin{figure*}
\centerline{\vbox{
\includegraphics[width=\textwidth]{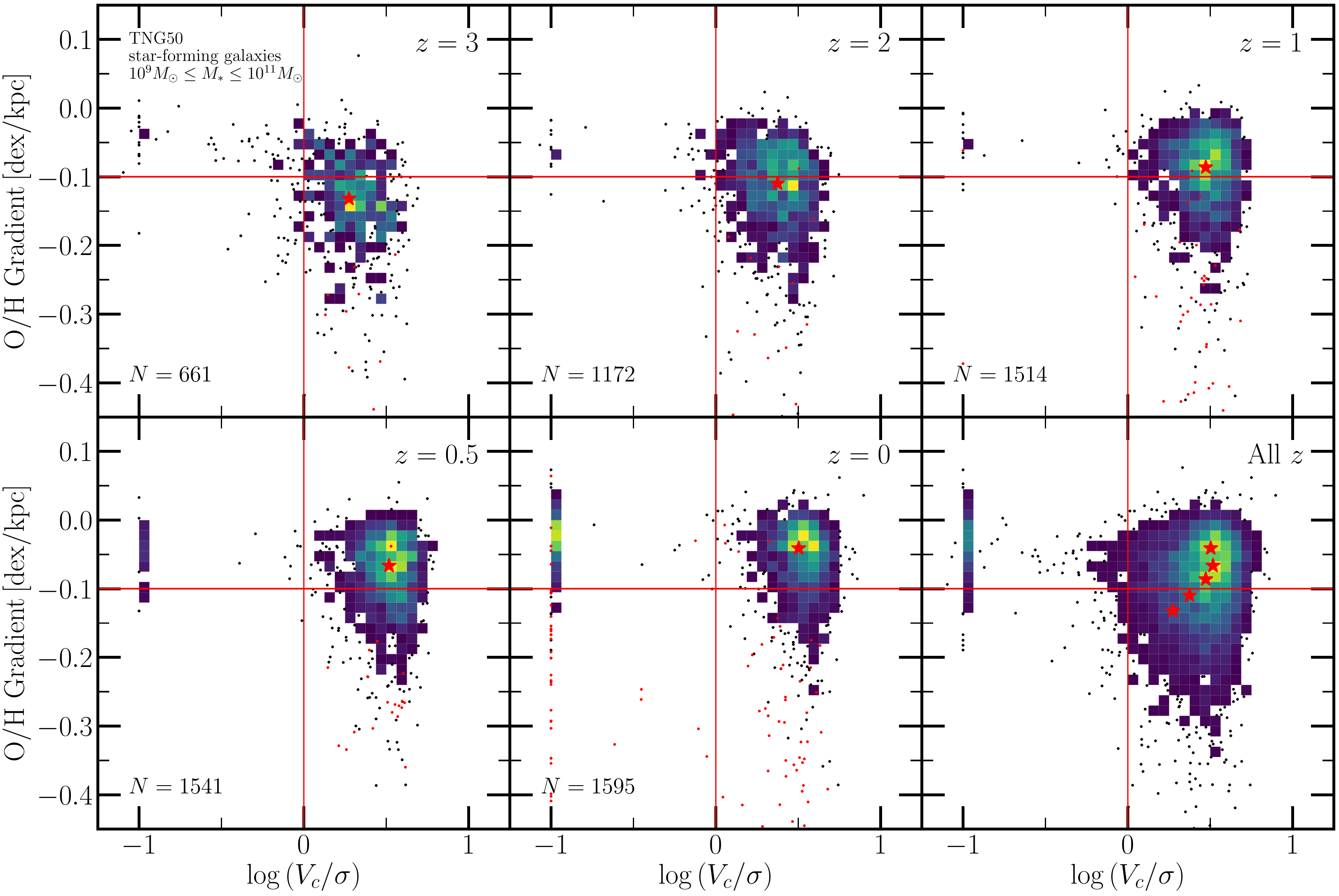}
}}
\caption{
The evolution through cosmic time of the relation between metallicity gradient ($\alpha$) and the galactic rotational-support parameter ($V_c / \sigma$, see Sections~\ref{sec:galkinematics}~and~\ref{sec:profgrad}). 
Each panel displays a gradient--$\log \left( V_c / \sigma \right)$ hybrid scatter plot/2D histogram of all galaxies in the selected TNG50 sample (see Section~\ref{sec:galselection}) at the redshift noted in each upper-right corner. 
Colored pixels are shown only for regions containing more than 3 galaxies -- otherwise, individual data points are shown. 
Red points show galaxies with unreliable gradients (gradient uncertainty $>0.1$ dex/kpc), and are therefore excluded from the all-$z$ panel. 
Typically, galaxies with $V_c / \sigma > 1$ are accepted as rotationally-supported. 
Vertical horizontal red lines mark this cutoff for rotationally-supported galaxies. 
Horizontal red lines divide the sample into those with steep metallicity gradients ($\alpha < -0.1 \, \mathrm{dex/kpc}$) and those with shallow/flat metallicity gradients ($\alpha > -0.1 \, \mathrm{dex/kpc}$). 
Red stars mark the median gradient and median $\log \left( V_c / \sigma \right)$ at each redshift. 
A discussion of this figure is given in Section~\ref{sec:gradkine}. 
}
\label{fig:kine}
\end{figure*}

In this section, we examine how metallicity gradients relate to galaxy kinematics (specifically, the kinematic parameter $V_c / \sigma$). 
Briefly, $V_c/\sigma$ is a ratio that compares the kinematic dispersion of a galaxy ($\sigma$) to its rotation ($V_c$). 
$V_c/\sigma < 1$ implies that the kinematics of a galaxy are dispersion-dominated, while $V_c/\sigma > 1$ implies they are rotation-dominated (i.e., the galaxy is rotationally-supported). 
Generally, rotationally-supported galaxies possess well-formed disks. 
The process by which the $V_c$ and $\sigma$ of a galaxy are determined is detailed in Section~\ref{sec:galkinematics}. 
Following \citetalias{Ma17}, for galaxies with velocity curves that do not meet the fit requirements of Section~\ref{sec:galkinematics}, we set $V_c/\sigma = 0.1$ (as they are dispersion-dominated). 
Some previous simulations and observations \citep[e.g.][]{Jones13, Leethochawalit16, Ma17} have found a significant positive correlation between $V_c/\sigma$ and gradient steepness. 
Such a correlation is intuitive, as metal-rich gas in dispersion-dominated galaxies will be more effectively radially redistributed, possibly serving to flatten metallicity gradients. 
Dispersion-dominated galaxies can be a result of mergers, or can simply be newly formed (see \citealt{Pillepich19}). 

Each panel of Figure~\ref{fig:kine} compares metallicity gradient and $V_c/\sigma$ for all galaxies at the redshifts noted in each upper-right corner. 
The bottom-right (``all-$z$'') panel shows data from all redshifts combined. 
The panels are further divided into four regions by red lines indicating the thresholds for rotationally-supported galaxies ($V_c / \sigma = 1$) and for steep metallicity gradients ($\alpha = -0.1$ dex/kpc). 
Thus, starting from the upper-left region of each panel and moving clockwise, each region contains galaxies that are (i) dispersion-dominated with flat gradients, (ii) rotation-dominated with flat gradients, (iii) rotation-dominated with steep gradients, and (iv) dispersion-dominated with steep gradients. 
The median gradient and $V_c/\sigma$ of each panel are marked by red stars. 
Individual red data points show galaxies with gradients that are highly uncertain (uncertainty $>0.1$ dex/kpc) -- these gradients are excluded from the all-$z$ panel. 

The progression of red stars in the all-$z$ panel shows that, with decreasing redshift, TNG50 star-forming galaxies evolve from being less to more rotationally supported, and their gradients from being steep to shallow. 
As should be expected, redshifts $z = $ 3 and 2 exhibit the largest fractions of dispersion-dominated galaxies. 
Most of these dispersion-dominated galaxies posses shallow gradients, although a few present moderately steep gradients -- however, the gradients of rotation-dominated galaxies at these redshifts are significantly steeper. 
Moreover, most of the dispersion-dominated galaxies that exhibit steep gradients lie near the threshold of being rotation-dominated. 
Most galaxies at these high redshifts (in our star-forming, centrals-only sample) are rotation-dominated (see \citealt{Pillepich19} for a full analysis), and most of these galaxies exhibit very strong gradients. 
Rotation-dominated galaxies at redshifts $z = $ 1 and 0.5 are approximately equally split between exhibiting shallow gradients and steep gradients. 
Star-forming dispersion-dominated galaxies at these redshifts are very rare, and those that exist possess shallow gradients. 
By redshift $z = 0$, the gradients of rotation-dominated galaxies are mostly flat. 
Also at this redshift, a significant sub-population of dispersion-dominated galaxies reappears -- so many that the median is noticeably pulled back towards dispersion dominance. 
This is likely a result of TNG50 AGN feedback, which significantly disrupts gas disks and gradients. 
Many of these disks are so disrupted that our methodology is not sufficient to reliably measure gradients.
While most gradients in the redshift $z = 0$ dispersion-dominated systems are shallow, some are steep. 
However, these steep gradients are highly uncertain in general, suggesting that they are mostly artifacts. 

We further discuss Figure~\ref{fig:kine} and TNG50 galaxy kinematics in Section~\ref{sec:dkine}.

\section{Discussion}

\label{sec:discussion}

\subsection{TNG50 metallicity gradients vs. simulations}
\label{sec:dsim}

\subsubsection{Metallicity gradient evolution}
\label{sec:dgradz}

\citetalias{Gibson13} showed that metallicity gradient redshift evolution could depend upon the strength of feedback implemented in galaxy formation models. 
Specifically, \citetalias{Gibson13} compared the metallicity gradient evolution of the MUGS and MaGICC simulations. 
Both MUGS and MaGICC rely on subgrid physics to model supernova (SN) feedback -- specifically, the adiabatic blastwave model \citep{Stinson06}. 
In this model, heat is injected into gas surrounding star particles. 
To prevent this energy from being quickly radiated away due to the high densities of star-forming environments, cooling is disabled within $\sim 100$ pc of the supernovae for a period of $\sim 10$ Myr, with exact values depending on ambient density and pressure \citep[][]{Brook12b, Gibson13}. 
The thermal energy injected by MUGS per SN is a factor of $\sim$2 less than that injected by MaGICC. 
Moreover, MaGICC includes radiation energy feedback while MUGS does not. 
As noted in \citetalias{Gibson13}, the weaker MUGS feedback model produces very steep metallicity gradients at high redshift while the stronger MaGICC model produces flat high-redshift gradients. 
This disparity follows naturally from stronger feedback driving stronger outflows that radially mix and redistribute enriched gas, thereby producing flatter metallicity gradients. 
At low redshift, the metallicity gradients produced by both feedback models are consistent with each other. 

\citetalias{Gibson13} was careful to use two simulations that differed negligibly besides the scaling of their feedback implementations. 
In this case, it is clear which feedback implementation is stronger and which is weaker. 
Moreover, \citetalias{Gibson13} ran the MaGICC and MUGS simulations on the same two galaxies (g1546 and g15784 from the MUGS simulation suite). 
By holding these variables constant, \citetalias{Gibson13} isolated the effects of feedback and was able to make conclusive statements regarding the feedback-dependence of gradients. 
In general, however, it is difficult to rank the strength of feedback between simulations that differ significantly overall (e.g. MUGS/MaGICC and TNG50). 
To model SN feedback, TNG50 relies on subgrid methods that are fundamentally different than the blastwave model of MUGS/MaGICC. 
TNG50 injects kinetic energy into local gas that is then hydrodynamically decoupled, thereby creating outflows. 
In addition, the effective equation of state (eEOS) of TNG50 includes ISM pressurization that might not otherwise be resolved \citep{Springel03}. 
Given these differences, it is not immediately clear how the strength of the TNG50 feedback implementation as a whole compares to that of MUGS or MaGICC. 
However, previous analyses of star formation and feedback in Illustris/TNG and FIRE may provide some insight into the differing high-redshift gradients of TNG and FIRE. 

As opposed to MUGS/MaGICC and TNG50, FIRE explicitly models stellar feedback. 
Several studies \citep[][]{Hopkins14, Muratov15, Sparre17} show that this explicit treatment leads to bursty star formation and correspondingly bursty feedback at high redshifts. 
\citetalias{Ma17} specifically demonstrates that this bursty high-redshift feedback drives powerful outflows that severely disrupt gas disks and their metallicity gradients. 
\citet{Faucher-Giguere18} suggests that all FIRE galaxies display bursty SF/feedback at high redshift because (i) stars only form in a few massive gravitationally-bound clouds (GBCs) and (ii) explicit stellar feedback reacts too slowly to halt the collapse of these clouds. 
High-mass FIRE galaxies eventually develop enough star-forming GBCs to achieve more stable star formation at low redshifts, allowing the build-up of steep gradients. 
In contrast, low-mass FIRE galaxies do not develop enough star-forming GBCs to stabilize and, resultantly, continue experiencing bursts of feedback that repeatedly erase gradients. 
On the other hand, the Illustris/TNG eEOS and kinetic wind model produce more-steady star formation and stellar feedback at both high and low redshifts, for both high- and low-mass galaxies \citep[][]{Sparre15, Torrey18, Torrey19}. 
In fact, galactic SFRs and metallicities in star-forming TNG galaxies have been demonstrated to vary smoothly on halo free-fall timescales \citep[][]{Torrey18, Torrey19}. 
Owing to this more-steady stellar feedback, we do not observe gas disk/gradient disruptions in star-forming galaxies on the same magnitude as those of FIRE. 
Consequently, TNG50 galaxies are able to form inside-out, thereby developing steeper high-redshift gradients -- the correlations between gradients and galaxy size/mass displayed in Figures~\ref{fig:mstar}~and~\ref{fig:corr} provide strong evidence for this inside-out formation \citep[e.g.][]{PB00}. 
Figures~\ref{fig:gibssim}~and~\ref{fig:mstar} show that high-mass FIRE galaxies -- the galaxies that least experience bursty feedback -- are the only star-forming FIRE galaxies that consistently exhibit gradients as steep or steeper than TNG50 gradients. 
In TNG50, such high-mass galaxies are also subject to kinetic AGN feedback, which further flattens their gradients beyond those of FIRE. 
As expected based on this discussion, Figures~\ref{fig:gibssim}~and~\ref{fig:mstar} also show that most low-to-intermediate mass star-forming FIRE galaxies exhibit significantly flatter gradients than those of TNG50, both at high and low redshift. 
We emphasize that neither the FIRE feedback model nor the TNG model is necessarily ``stronger'' in the same way that MaGICC is stronger than MUGS. 
However, the FIRE stellar feedback model is definitively more bursty and disruptive to gradients than that of TNG. 
Our results, combined with those of \citetalias{Gibson13} and \citetalias{Ma17}, may suggest that high-redshift gradients depend on both the strength of feedback \textit{and} the timescales over which this feedback acts. 

\subsubsection{Galaxy kinematics}
\label{sec:dkine}

Figure~\ref{fig:kine} shows that our TNG50 sample predominantly contains rotation-dominated galaxies. 
This bias is likely a result of our selection for central-only, star-forming galaxies. 
Star formation requires gas to cool and coalesce into dense clumps, and these clumps typically form in galactic disks. 
Galactic disks, in turn, necessitate some degree of rotational support. 
Still, we observe significant sub-populations of dispersion-dominated galaxies at appropriate redshifts -- specifically, redshifts $z = $ 3 and 2 due to mergers and disk formation, and redshift $z = 0$ due to AGN feedback. 

Generally, Figure~\ref{fig:kine} illustrates that, as time progresses, (i) metallicity gradients become more flat and (ii) galaxies become more disky. 
The latter fits well with the current conception of galaxy formation and evolution (see \citealt{Genel18} and \citealt{Pillepich19} for extended discussions of stellar and gas disk evolution, respectively). 
The former is likely motivated by several factors. 
Section~\ref{sec:gradr} demonstrates that TNG50 metallicity gradients are strongly inversely correlated with \RSFR, which is effectively a measure of disk size. 
Thus, as TNG50 disks accrete mass and grow, their gradients continuously level out -- this smooth growth and leveling with redshift is shown in Figure~\ref{fig:corr}, and may be a product of the TNG50 stellar feedback model. 
Specifically, the decoupled winds and effective equation of state of this feedback model allow disks in our selected mass range to remain relatively stable with time, until kinetic AGN feedback eventually overwhelms disks. 
TNG50 AGN in a low-accretion state produce powerful kinetic winds that are \textit{not} hydrodynamically decoupled. 
Thus, unlike TNG50 stellar feedback, this kinetic AGN feedback can, with time, significantly disrupt gas disks and flatten gradients \citep[][]{Weinberger17, Pillepich18b, Nelson18}. 
Still, for our $10^{9} M_{\odot} < M_* < 10^{11} M_{\odot}$ star-forming galaxy sample, AGN feedback is most disruptive (i) in high-mass galaxies and (ii) at late times \citep[][]{Weinberger18, Zinger20}, although it can be active in some capacity as early as redshift $z \sim$ 2--4 \citep[][]{Donnari19, Nelson19b}. 
We therefore conclude that the gradual flattening of metallicity gradients is primarily a result of the TNG50 stellar feedback model, although kinetic AGN feedback may also contribute significantly. 

Although we observe a few cases of steep gradients in dispersion-dominated galaxies, several caveats exist for these cases. 
In almost every example, these cases include galaxies that are near the rotation-dominated threshold, gradients that are near the shallow threshold, and/or gradients that are highly uncertain. 
Still, steep gradients in dispersion-dominated galaxies are not impossible. 
For example, \citet{Rupke10b} shows that gradients of merging galaxies require some time to flatten after kinematic perturbations from interactions. 
The same may be true for kinematic perturbations via feedback. 
Thus, steep gradients may still be observable in the early stages of these disruptive events. 

As expected based on the results of previous simulations \citepalias[e.g.][]{Ma17} and observations \citep[e.g.][]{Jones13, Leethochawalit16}, we find that steep gradients are almost exclusively found in rotation-dominated galaxies, and that dispersion-dominated galaxies exhibit almost exclusively shallow gradients. 
We also find a significant sub-population of rotation-dominated galaxies with shallow gradients -- this sub-population is in the minority at early times and the majority at late times. 
Overall, we judge our results to be qualitatively consistent with the gradient-kinematics relationships of other simulations and observations. 

\subsection{TNG50 metallicity gradients vs. observations}
\label{sec:dobs}

\subsubsection{Size correlation and metallicity gradient evolution}
\label{sec:dgradrobs}

\begin{figure*}
\centerline{\vbox{
\includegraphics[width=\textwidth]{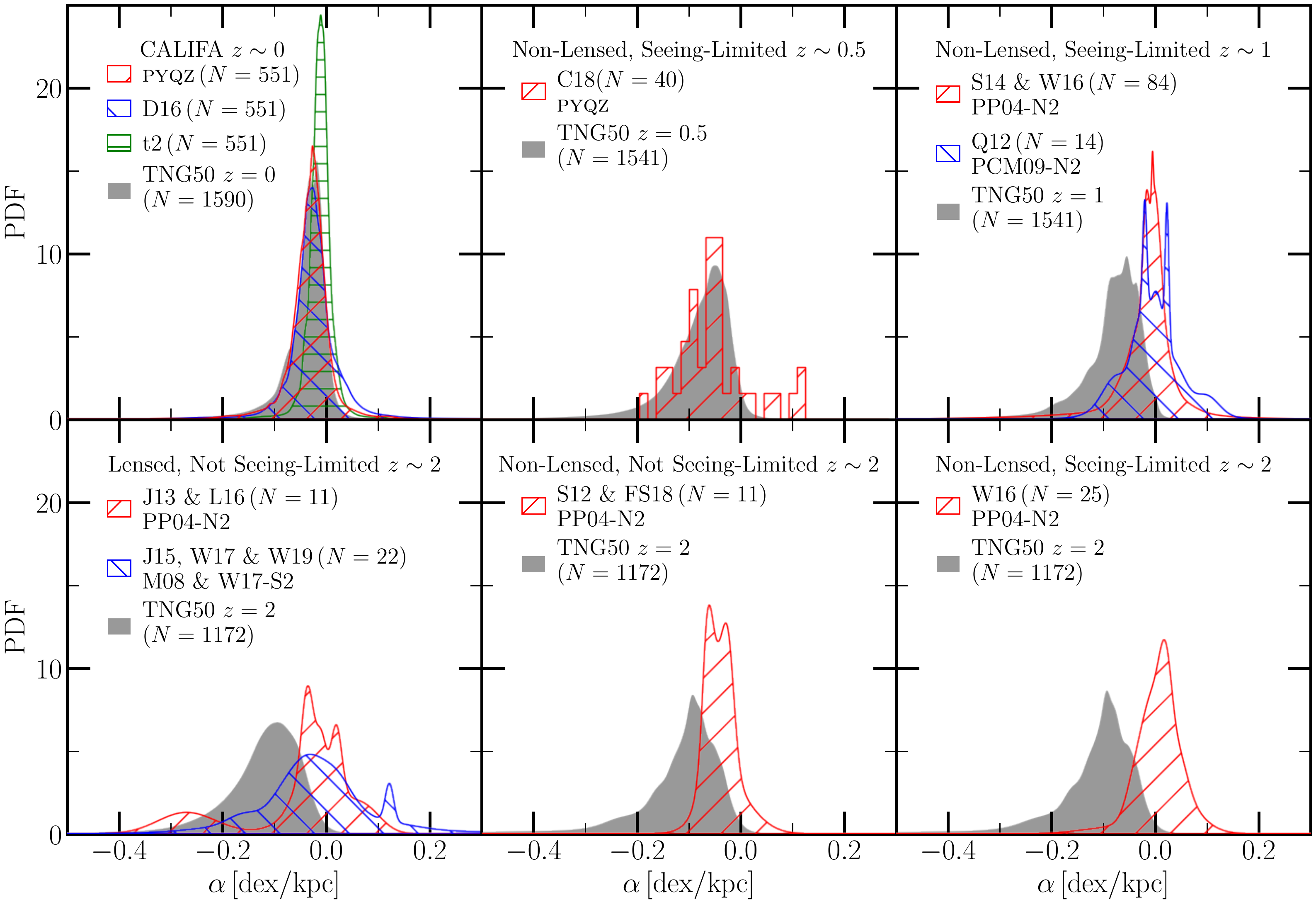}
}}
\caption{
TNG50 star-forming galaxy gradient distributions v. observed gradient distributions from redshift $z=$0--3. 
We compare TNG50 gradient distributions to that of CALIFA \citep[][]{Sanchez12b, Sanchez14, Sanchez-Menguiano16} at redshift $z = 0$, that of \citet{Carton18} at redshift $z \sim 0.5$, those of \citet{Queyrel12}, \citet{Stott14}, and \citet{Wuyts16} at $z \sim 1$, and those of \citet{Swinbank12}, \citet{Jones13}, \citet{Jones15}, \citet{Leethochawalit16}, \citet{Wuyts16}, \citet{Wang17}, and \citet{Wang19} at $z \sim 2$. 
In all cases, we scale the TNG50 gradient distribution to compensate for differences in mass distribution between the TNG50 sample and the observed samples. 
Further, we sort studies into several groups based on (i) whether the galaxy samples are lensed or non-lensed, (ii) whether the observations are or are not seeing-limited, and (iii) the indicator(s)/calibrator(s) used to infer abundances. 
The top of each panel indicates which category of (i) and (ii) the observations displayed fall into, and the indicator(s)/calibrator(s) for each distribution are given in the legend. 
The indicators/calibrators used for these data are \textsc{pyqz} \citep[][]{Dopita13}, D16 \citep[][]{Dopita16}, t2 \citep[][]{Pena-Guerrero12}, PP04-N2 \citep[][]{PP04}, PCM09-N2 \citep[][]{Perez-Montero09}, M08 \citep[][]{Maiolino08}, and W17-S2 \citep[][]{Wang17}.
}
\label{fig:lspc}
\end{figure*}

TNG50 makes clear predictions about the redshift evolution of metallicity gradients -- specifically, that these gradients evolve in lockstep with the size of galaxies hosting them. 
Recent results from the EAGLE simulation \citep[][]{Tissera19} also predict a correlation between disk size (specifically half-mass gas radius) and gradient at redshift $z = 0$. 

Several observations support the predictions of TNG50 and EAGLE at redshift $z = 0$, finding that metallicity gradients of local galaxies do indeed scale with disk size. 
Early on, \citet{Diaz89} and \citet{VCE92} noted that scatter among metallicity gradient measurements could be decreased via normalizing gradients by disk scale lengths. 
Soon after, \citet{Zaritsky94} and \citet{Garnett97} found that this normalization eliminated correlations between gradient and both galaxy luminosity and Hubble Type, hypothesizing that galaxies ``are homologous with regard to chemical evolution'' and form inside-out. 
This homologous hypothesis was further supported by analytic models of chemical evolution \citep[e.g.][]{PB00}, and was later confirmed when two large-sample IFU studies, \citet{Sanchez12b} and \citet{Sanchez14} (CALIFA, see \citealt{Sanchez12a}), found a characteristic metallicity profile and gradient among 300+ local galaxies after normalizing by effective disk radius. 
With further confirmation from several subsequent studies including \citet{BK15}, \citet{Ho15}, \citet{Sanchez-Menguiano16}, \citet{Sanchez-Menguiano18}, there is now a broad consensus that a common metallicity profile and gradient exists at redshift $z = 0$. 
This common metallicity profile and gradient suggest that local galaxies formed inside-out \citep[][]{Sanchez14, Sanchez-Menguiano16}. 
Still, some large-sample studies of the local Universe \citep[e.g. MaNGA;][]{Belfiore17} do not find a correlation between galaxy size and gradient. 

Given the existence of a characteristic gradient in the local Universe, one may expect a similar scaling between gradient and galaxy size at higher redshifts. 
Because disk scale lengths decrease with redshift, this scaling should lead to steeper gradients at higher redshift. 
However, most observations of more distant galaxies at redshifts $z \gtrsim 1$ measure almost exclusively shallow, flat, or inverted gradients that do not scale with galaxy size. 
The disparity between gradients measured at redshift $z \sim 0$ and those measured at higher redshifts is emphasized in Figure~\ref{fig:lspc}, which compares the gradients of TNG50 (at appropriate redshifts) to those measured at redshift $z \sim 0$ by CALIFA \citep[][]{Sanchez12b, Sanchez14, Sanchez-Menguiano16}, redshift $z \sim 0.5$ by \citet{Carton18}, redshift $z \sim 1$ by \citet{Queyrel12}, \citet{Stott14}, and \citet{Wuyts16}, and redshift $z \sim 2$ by \citet{Swinbank12}, \citet{Jones13}, \citet{Jones15}, \citet{Wuyts16}, \citet{Leethochawalit16}, \citet{ForsterSchreiber18}, \citet{Wang17}, and \citet{Wang19}. 
A significant fraction of observed gradients come from galaxies outside the $10^{9}\,M_{\odot} \leq M_* \leq 10^{11}\,M_{\odot}$ mass range that we have sampled in TNG50. 
These gradients are excluded from Figure~\ref{fig:lspc} to allow an even-handed comparison with our selected TNG50 sample. 
For the same reason, we scale each TNG50 gradient distribution to address differences between the mass distribution of our TNG50 sample and that of observations. 
Moreover, because (i) different calibrators are known to return different abundance inferences (and therefore gradients) and (ii) observations with varying spatial resolution measure different gradients, we only group studies that (i) use the same calibrators and (ii) make observations with similar spatial resolutions. 
Addressing point (ii), we separate references (and their measurements) into four spatial resolution categories, reasoning that their measurements can be from either lensed (L) or non-lensed (NL) galaxies via observations that are either seeing-limited (S) or not seeing-limited (NS). 
For comparison, our analysis utilizes a spatial resolution of 100 pc, as described in Section~\ref{sec:profgrad}. 
The nearby (NL-S, with spatial resolutions of $\sim$ 1 kpc) gradients measured by CALIFA (via the \textsc{pyqz} and D16 calibrators) are mostly negative, forming an approximate log-normal distribution that agrees exceptionally well with the redshift $z = 0$ gradient distribution of TNG50. 
Also, the gradients of these studies exhibit the same gradient-size scaling found among TNG50 galaxies. 
However, we caution that the CALIFA gradients inferred via other calibrators (e.g. t2) disagree significantly with both the \textsc{pyqz}/D16 CALIFA gradients and the redshift $z = 0$ TNG50 gradients. 
The more distant (NL-S, with spatial resolution $\sim$ 4 kpc) gradients measured by \citet{Carton18} agree reasonably well with the redshift $z = 0.5$ TNG50 gradient distribution, although this study observes a more significant sub-population of positive gradients than TNG50. 
However, \citet{Carton18} does not claim to find evidence for a correlation between galaxy size and gradient. 
The $z \approx 1$ (NL-S, with spatial resolutions of $\sim$ 5--6 kpc) gradients measured by \citet{Queyrel12}, \citet{Stott14}, and \citet{Wuyts16} disagree significantly with the redshift $z = 1$ TNG50 gradient distribution. 
Compared to the TNG50 gradient distribution, the distributions of these studies are decidedly more Gaussian and exhibit a relative over-abundance of flat and positive gradients. 
Moreover, these studies do not observe a gradient-size correlation. 
It is important to note that the gradients measured by most large-sample IFU studies at $z \gtrsim 1$ are mostly consistent with those measured by \citet{Wuyts16}, with normally distributed gradients that are predominantly shallow-to-flat and no apparent correlation between galaxy size and gradient. 
However, the same is not true for the high-resolution redshift $z \sim 2$ L-NS gradients of \citet{Jones13}, \citet{Jones15}, \citet{Leethochawalit16}, \citet{Wang17}, and \citet{Wang19}. 
These studies boast sub-kpc spatial resolution down to 200--300 pc. 
While still measuring some flat and inverted gradients, these studies find significantly more steep gradients than any other high-redshift studies with lesser spatial resolution. 
As high-redshift observations move towards lower spatial resolutions, they appear to disagree progressively more with the redshift $z = 2$ TNG50 distribution. 
While redshift $z \sim 2$ NL-NS studies \citep[][]{Swinbank12, ForsterSchreiber18} (with $\sim$ 1--2 kpc spatial resolution) measure predominantly negative gradients, these gradients are not quite steep enough to match the TNG50 distribution. 
\citet{Wuyts16}, the redshift NL-S $z \sim 2$ study (with $\sim$ 5 kpc spatial resolution), measures predominantly positive gradients that disagree even more significantly with the TNG50 distribution. 
These disagreements at high-redshift may point to possible limitations of the TNG model, and/or to potential systematic errors in observations -- we discuss these matters in Section~\ref{sec:dtension}. 

\subsubsection{Stellar mass correlation}
\label{sec:dmobs}

As described in Section~\ref{sec:gradm}, we observe a weak positive correlation between galaxy stellar mass and metallicity gradient measured on physical scales. 
Some studies of nearby galaxies \citep[e.g.][]{Sanchez12b, Sanchez14, Ho15, Sanchez-Menguiano16, Sanchez-Menguiano18} find a similar positive correlation, although this correlation seems to disappear at higher redshifts. 
Figure~\ref{fig:mstar} displays the TNG50 gradient-$M_*$ relation compared to that of the redshift $z \sim 0$ \citet{Ho15} sample (which includes supplemental gradient measurements from \citealt{Rupke10b} and \citealt{Sanchez12b}) along with the CALIFA sample \citep[][]{Sanchez12b, Sanchez14, Sanchez-Menguiano16}, the $z \sim 0.5$ \citet{Carton18} sample, the $z \sim 1$ \citet{Stott14} and \citet{Wuyts16} NL-S sample, and the $z \sim 2$ L-NS sample \citep[][]{Jones13, Jones15, Leethochawalit16, Wang17, Wang19}. 

The gradients of \citet{Ho15}, CALIFA, and TNG50 are compared in the redshift $z = 0$ panel of Figure~\ref{fig:mstar}. 
The (\textsc{pyqz}) CALIFA and \citet{Ho15} data are shown as orange and magenta points, respectively. 
One will note that the entire \citet{Ho15} sample and most of the CALIFA sample lies within the spread of the TNG50 gradient--$M_*$ distribution. 
Observers (e.g. \citealt{Sanchez12b}; \citealt{Sanchez14}; \citealt{Ho15}) typically normalize gradients by either $R_{25}$ (the radius of the $25 \, \mathrm{mag/arcsec^2}$ B-band isophote) or $R_e$ (the half-flux ``effective-radius'') and find that these normalized gradients no longer correlate with $M_*$. 
Because we measure a different galaxy radius (\RSFR) based on the radial distribution of star formation rather than flux, we do not directly compare our normalized gradients to those of observations. 
However, we note that TNG50 gradients normalized by \RSFR\, also do not correlate with $M_*$ (see Section~\ref{sec:gradm}~and~Figure~\ref{fig:mstar}). 
Thus, TNG50 gradients at redshift $z = 0$ seem to accurately characterize gradients observed locally. 
Still, we caution that some large-sample studies of nearby galaxies (e.g. \citealt{Belfiore17}) find a weak positive correlation between normalized gradient steepness and stellar mass. 

The gradients of \citet{Carton18} and TNG50 are compared in the redshift $z = 0.5$ panel of Figure~\ref{fig:mstar}. 
Except for a few outlying positive gradients, the measurements of \citet{Carton18} appear to mostly agree with TNG50 gradients at $z = 0.5$. 
However, \citet{Carton18} finds evidence for a positive correlation between normalized gradient steepness and stellar mass. 
No similar correlation is found among normalized TNG50 gradients. 
The following two panels compare the redshift $z = $ 1 and 2 TNG50 gradient-$M_*$ distribution to observations, showing siginificant disagreements. 
Most of these studies measure a relative overabundance at flat and positive gradients. 
TNG50 is in best agreement with the highest-resolution (L-NS) observations. 
Moreover, none of these studies at redshifts $z \geq 1$ show a significant correlation between galaxy stellar mass and gradient. 

Given the notable qualitative and quantitative agreements shown in Figures~\ref{fig:mstar}~and~\ref{fig:lspc} at redshift $z = 0$, we conclude that TNG50 models gradients in the local Universe with great accuracy. 
Even at redshift $z = 0.5$, the TNG50 gradients seem to mostly agree with observations in both Figures~\ref{fig:mstar}~and~\ref{fig:lspc}. 
However, at redshift $z = 1$ and beyond, TNG50 predicts much steeper gradients than those observed by most large-sample studies and does not capture the lack of gradient correlations. 

Figure~\ref{fig:mstar} also includes gradient--$M_*$ data from several other simulations -- \citet{Tissera16}, FIRE \citep{Ma17}, and EAGLE \citep{Tissera19}. 
We find significant agreement between the gradient--$M_*$ relations of TNG50 and \citet{Tissera16} at redshift $z = 0$, although this agreement lessens at higher redshifts. 
However, the flat gradients of low-mass galaxies and steep gradients of high-mass galaxies observed in FIRE (at all redshifts) are not present in TNG50, nor are the (typically) flat gradients at all masses observed in EAGLE at redshift $z = 0$. 
Still, \citet{Tissera19} observes a gradient--$M_*$ relation similar to that of TNG50 among galaxies with quiet merger histories. 

\subsubsection{Apparent tension with high-redshift observations}
\label{sec:dtension}

The origin of the disagreements between observed gradients/correlations and those of TNG50 is not immediately clear. 
It is possible that issues exist for both the current modelling approach and the current abundance measurement methodology. 
In this subsection, we discuss some potential issues that could lead to either inaccurate simulated gradients on the part of TNG50 or inaccurate measured gradients on the part of observations. 
For this discussion, we assume that any discrepancies between our simulation and observations are not the result of our measurement choices. 
All simulated gradients in this paper are from mass-weighted metallicity profiles of galaxies projected face-on, which does not match the methodology of observers. 
We postpone addressing the impact of these methodological differences to a future work. 

\subsubsection*{On the TNG ISM and stellar feedback models}

Although the TNG simulations attempt to accurately capture many properties of our Universe, some necessary simplifications may make the model insufficient to fully reproduce the complexity of metallicity gradients, their correlations, and their evolution. 
While virtually any aspect of the TNG model could influence the radial distribution of metals in galaxies, we identify two potentially significant sources of gradient inaccuracies -- the implementations of (i) the ISM and (ii) galactic winds. 
A more detailed discussion of these implementations is given in Section~\ref{sec:tng}. 

TNG does not explicitly model some small-scale (i.e., $<100 \, \mathrm{pc}$ -- on the order of GMCs) phenomena (e.g. turbulence) that contribute to ISM pressurization, and instead models this pressurization via a two-phase, effective equation of state model described in \citet{Springel03}. 
Still, it is well-known that turbulence provides a significant component of ISM pressure support. 
Likely, this is especially true at higher redshifts where high dispersions in gas velocity are commonly observed. 
While \citet{Pillepich19} demonstrates that the larger-scale turbulence in TNG50 gas disks agrees reasonably well with observations (both at high-redshift and as a function of redshift), unresolved small-scale turbulence could still serve to radially mix chemically enriched gas and flatten metallicity gradients. 
Should such turbulence (or any other small-scale phenomena) prove to play a significant role in redistributing metals, TNG50 may be unable to capture this effect. 

Another phenomenon by which metals can be radially redistributed is galactic winds. 
Without a doubt, varying the TNG properties/parameters of wind energy, velocity, mass loading, metal loading, and/or recoupling could significantly impact the radial distribution of chemically enriched gas, which would necessarily alter metallicity gradients. 
For example, \citet{Grand19} demonstrates that varying the wind metal loading factor in the Auriga simulations \citep[][]{Grand17} significantly flattens metallicity gradients. 
Additionally, should winds exhibit a different scaling with redshift than that featured in the TNG model, this could change the tension between TNG50 and observations as a function of redshift. 

Because the GMC-scale properties of the ISM are not thoroughly resolved in TNG50, probing its metallicity gradients represents a strict test of the simulation that pushes the limits of its resolution, physics, and feedback implementation. 
Still, this test provides an important comparison to the results of previous simulations and observations, is predictive, and may prove accurate should observed metallicity gradients depend mostly on the larger-scale phenomena modelled in TNG. 

\subsubsection*{On observational systematics}

\begin{figure}
\centerline{\vbox{\hbox{
\includegraphics[width=0.45\textwidth]{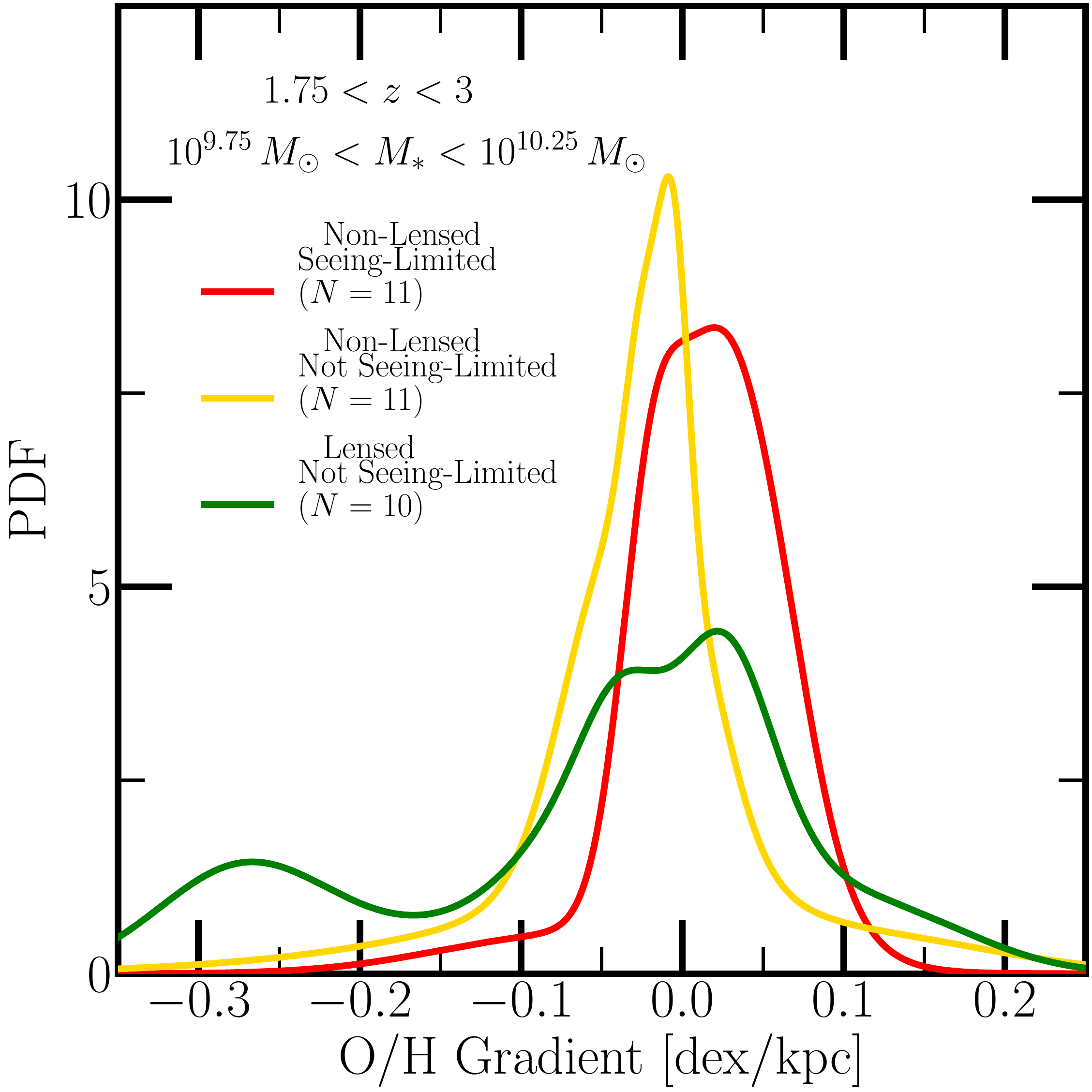} 
}}}
\caption{
The current body of observed metallicity gradients at redshifts $1.8 < z < 3$. 
Each curve shows the PDF of measurements from either lensed (L) or non-lensed (NL) galaxies via observations that were either seeing-limited (S) or not seeing-limited (NS). 
L-NS observations ($N = 10$) include those of \citet{Jones10}, \citet{Jones13}, \citet{Leethochawalit16}, \citet{Wang17}, and \citet{Wang19}, NL-NS ($N = 11$) those of \citet{Swinbank12}, \citet{Molina17}, and \citet{ForsterSchreiber18},and NL-S those of \citet{Wuyts16}. 
We include only galaxies with stellar mass within $1\sigma$ of $10^{9.75} \, M_{\odot} \leq M_* \leq 10^{10.25} \, M_{\odot}$, and model measurements with asymmetric uncertainties as two combined half-Gaussians. 
We note that not all studies use the same indicator(s)/calibrator(s).
}
\label{fig:hiz}
\end{figure}

On the other hand, the disagreements between TNG50 and observations at higher redshifts (and between observations themselves at lower redshifts) could potentially be the result of systematics in metallicity gradient measurements. 
Many sources of possible systematics have been proposed, including but not limited to (i) the metallicity diagnostic calibrations themselves, (ii) the signal to noise ratio (SNR) of observations and, (iii) the angular resolution and binning of observations. 
Thus, these high-redshift gradient disagreements may occur because our methodology does not attempt to mock the methods of observations and their systematics. 

The effects of decreasing angular resolution and SNR have been thoroughly investigated by several studies~\citep[e.g.][]{Yuan13, Mast14, Poetrodjojo19, Acharyya20}. 
\citet{Yuan13}, \citet{Mast14}, and \citet{Poetrodjojo19} artificially downgrade the quality of high-resolution, high-SNR observations to determine that decreasing angular resolution, bin count, and/or SNR significantly flattens measured metallicity gradients. 
These results are confirmed and emphasized by \citet{Acharyya20}, which utilizes isolated galaxy simulations to generate mock data. 
This novel approach allows the study to fully disentangle the effects of varying SNR, angular resolution, and bin count on measured metallicity gradients. 
Systematics introduced by angular resolution may explain why, even at redshift $z \approx 0$, lower-resolution IFS surveys (e.g. MaNGA, see \citealt{Belfiore17}) do not observe the galaxy size/mass--gradient correlations observed by higher-resolution IFS surveys (e.g. CALIFA) and in TNG50. 

In Figure~\ref{fig:hiz}, we sort the current body of high-redshift ($1.8 < z < 3$) gradient measurements of $10^{9.75} \, M_{\odot} < M_* < 10^{10.25} \, M_{\odot}$ galaxies into several angular resolution-based categories and compare their distributions to determine whether observations might show evidence for a systematic angular resolution bias.
We select the $10^{9.75} \, M_{\odot} < M_* < 10^{10.25} \, M_{\odot}$ because it contains several measurements from each category, allowing a fair comparison. 
The references placed into each category are given in the caption of Figure~\ref{fig:hiz}. 
The high-resolution category is the ``lensed, not seeing-limited'' (L-NS; sub-kpc spatial resolution) category, the intermediate-resolution category is the ``non-lensed, not seeing-limited'' (NL-NS; $\sim$ 1--2 kpc spatial resolution), and the low-resolution is the ``non-lensed, seeing-limited'' (NL-S; $\sim$ 5 kpc spatial resolution) category. 
Indeed, we find that the distributions of these two categories differ significantly. 
In fact, the L-NS distribution is drastically different from that of the other categories. 
The L-NS distribution is much wider than that of the lower-resolution distributions. 
This is consistent with the results of \citet{Yuan13}, \citet{Mast14}, \citet{Poetrodjojo19}, and \cite{Acharyya20}, suggesting that insufficient spatial resolution may indeed be systematically flattening high-redshift gradient measurements. 
However, we note that \citet{Wang17} and \citet{Wang19} -- two studies used to construct the L-NS distribution -- use different indicators/calibrators than the other studies included. 
This may contribute to the discrepancy between distributions. 

There are many potential reasons that gradients could vary significantly with angular resolution. 
The first and most obvious is beam-smeaing, which can radially redistribute emission line flux and thereby flatten gradients. 
It is also hypothesized that angular resolution affects measured gradients due to an observed overlap between individual \hii\, regions and (i) other \hii\, regions of differing physical properties and/or (ii) regions of diffuse ionized gas. 
Individual \hii\, regions are typically of order 10--100 pc -- at least an order of magnitude smaller than the approximately kpc scales (or worse) probed by most IFU surveys. 
Thus, typical spaxels may contain many individual \hii\, regions, each with its own physical conditions.
Because of the complex relationship between \hii\, region physical properties, strong line ratios, and calibrators, some specific \hii\, regions may dominate the derived metallicity inference \citep[][]{Kewley19}. 
Moreover, observations with poor angular resolution may be unable to isolate \hii\, regions from diffuse ionized gas (DIG). 
DIG has physical conditions that contrast significantly with those of \hii\, regions, invalidating several of the physical assumptions used to calibrate strong line metallicity diagnostics \citep[][]{Zhang17, Poetrodjojo19}. 
Thus, \hii\, region metallicities derived from emission lines with DIG contributions may be inaccurate. 

A similar issue may arise by attempting to measure gradients with poor spectral resolution. 
Models used to calibrate strong line diagnostics assume that the ionization in the \hii\, region is thermal. 
However, this assumption is invalidated if an \hii\, region is subject to shocks, which also ionize material and thereby contribute a non-thermal shock component to emission lines. 
Because material ionized by shocks moves at high speed, the shock components of the emission lines will be doppler broadened and can therefore be removed via sufficient spectral resolution and careful analysis. 
But without sufficient spectral resolution, this shock component could affect the strong line ratio measurement and contribute to inaccuracies in the abundance measurement \citep[][]{Rich11, Kewley13}.

The accuracy of gradient measurements is also dependent upon the accuracy of strong line diagnostic calibrations themselves. 
Strong line diagnostics are calibrated using both auroral line diagnostics and theoretical models of \hii\, regions. 
Because the former requires observations of faint auroral lines, only \hii\, regions in nearby galaxies can be used. 
Thus, strong line diagnostics are calibrated only to the environment at redshift $z \approx 0$. 
If any aspect of the high-redshift environment relevant to ionization in \hii\, regions -- e.g. ISM pressure and electron density, ionization parameter, or ionizing background radiation -- is different than that at low-redshift, strong line calibrations may be insufficient for high-redshift measurements \citep[][]{Carton18, Kewley19}. 
By combining photoionization models with observations of redshift $z \sim$ 2--3 galaxies, \citet{Strom17} and \citet{Strom18} find evidence that the redshift $z \sim 0$ calibrations may indeed be invalidated at high-redshift.
Perhaps most notably, they do not observe a significant correlation between \hii\, region metallicity and ionization parameter in these high-redshift galaxies, in stark constrast to the strong correlation observed between these parameters locally.
Moreover, theoretical strong line diagnostics rely on models of stellar population synthesis and photoionization. 
As \citet{Kewley19} describes, these models may have limitations that could potentially impact gradient measurements. 
Briefly, the current \hii\, region modelling paradigm would significantly benefit from (i) stellar tracks and opacity tables of significantly more metallicities to achieve high-resolution strong line diagnostics, (ii) quality observations of \hii\, regions to constrain their ionization, temperature, and density structure, along with their ionizing radiation field, (iii) quality observations of DIG to better model, diagnose, and remove its contamination, and (iv) extended model implementations of non-local thermodynamic equilibrium, scattering, stellar rotation, multiple star systems/clusters, non-ideal structure/geometries, and 3D radiative transfer \citep[][]{Kewley19}. 

The James Webb Space Telescope (JWST) along with its Near Infrared Spectrograph (NIRSpec) and Near-Infrared Imager and Slitless Spectrograph (NIRISS), as well as the Extremely Large Telescopes (ELTs; e.g. Thirty Meter Telescope, Giant Magellan Telescope, European Extremely Large Telescope) and their spectrographs will be capable of either validating or challenging the current body of high-redshift metallicity gradient measurements  \citep[][]{Yuan11, Wuyts16, Wang19, Maiolino19, Curti20b}. 
These telescopes and their instruments will have the wavelength sensitivity, angular/spectral resolution, and collecting area required to measure metallicity via auroral lines rather than via strong lines \citep[][]{Maiolino19, Curti20b}. 
These auroral line measurements will allow observers to (if needed) more accurately calibrate strong line metallicity diagnostics for the varying environments out to and at high-redshift. 
Moreover, the light-gathering power and angular resolution of these telescopes translate to high-SNR data at sub-kpc spatial resolution for redshifts $z \lesssim 3.5$ \citep[][]{Wuyts16, Wang19}. 
These factors combine to allow high-redshift gradient measurements via strong line diagnostics with miniminal contamination from all the potential systematics described above. 

\section{Summary and Conclusions}

From the TNG50 simulation, we select star-forming central galaxies in the stellar mass range $10^9 \, M_{\odot} < M_* < 10^{11} \, M_{\odot}$ with gas mass $M_{\mathrm{gas}} > 10^{9} \, M_{\odot}$ at redshifts $z =$ 3, 2, 1, 0.5, and 0 (Sections~\ref{sec:tng}~and~\ref{sec:galselection}). 
We rotate galaxies to a face-on orientation based on angular momentum-derived inclination angles and measure their kinematic properties via mock long-slit spectroscopy and a simple disk model (Sections~\ref{sec:galphysprops}~and~\ref{sec:galkinematics}). 
We define several galactocentric radii based on the galactic radial distributions of star formation, including \RSFR\, -- the 50\% total SFR radius. 
These radii are used to distinguish the ``star-forming region'' of each galaxy (Section~\ref{sec:galphysprops}). 
We map the spatially-resolved gas-phase, mass-weighted metal abundances of each galaxy and use these maps to construct radial metallicity profiles.
We then use these radial metallicity profiles to measure the metallicity gradient of each galaxy star-forming region (Section~\ref{sec:profgrad}; Figure~\ref{fig:subs}). 

We divide the sample into four stellar mass bins at each redshift and analyze the characteristics of metallicity profiles and gradients within and between bins (Section~\ref{sec:gradz0}). 
At all redshifts, we find that the metallicity gradient distribution of each stellar mass bin is well-characterized by a log-normal distribution (Section~\ref{sec:gradz0}; Figure~\ref{fig:proffit}). 
We fit the gradient distributions accordingly and present the redshift evolution of gradient distributions in Table~\ref{tab:gradvz} and Figure~\ref{fig:gibssim}. 
We compare the TNG50 gradient redshift evolution to the existing body of simulated and observed gradients in Figures~\ref{fig:gibssim}~and~\ref{fig:gibsobs}, respectively. 
We also examine how metallicity gradients relate to galaxy stellar mass (Section~\ref{sec:gradm}; Figure~\ref{fig:mstar}), galaxy size (Section~\ref{sec:gradr}; Figure~\ref{fig:corr}), and galaxy kinematics (Section~\ref{sec:gradkine}; Figure~\ref{fig:kine}), as well as how these relations evolve with redshift. 

Our primary conclusions from these analyses are as follows:

\begin{enumerate}
    \item{We find that metallicity profiles inside the star-forming region of TNG50 galaxies are well-approximated by an exponential (i.e., log-space linear) function with some metallicity gradient and intercept (Equation~\ref{eq:gradfit}).
    Outside the star-forming region, however, the metallicity gradient may change significantly (Figure~\ref{fig:subs}).}
    \item{Median metallicity profiles constructed from galaxies of similar stellar mass exhibit an offset in metallicity intercept that increases with stellar mass (an effect of the mass-metallicity relation). 
    Still, all median metallicity profiles share a common shape, especially if normalized by galaxy size. 
    The metallicity profiles of individual galaxies remain generally close to the median metallicity profile for their stellar mass, rarely deviating by more than $\sim 0.3$ dex (Figure~\ref{fig:med0}).}
    \item{TNG50 predicts a roughly monotonic increase in gradient steepness with redshift at a rate of approximately $-0.02 \left[ \mathrm{dex \, kpc^{-1}} / \Delta z \right]$, regardless of galaxy stellar mass.}
    \item{TNG50 high-redshift gradients are significantly steeper than those of FIRE, especially for low-to-intermediate mass galaxies (Figure~\ref{fig:gibssim}). 
    While TNG50 stellar feedback is not necessarily weaker than that of FIRE, it is definitively less bursty and disruptive (Section~\ref{sec:dsim}). 
    This result may suggest that high-redshift gradients are sensitive to the timescales over which feedback acts.}
    \item{TNG50 predicts a characteristic, redshift-invariant normalized gradient of approximately $-0.3$ dex/\RSFR\, in rotation-dominated galaxies (Figure~\ref{fig:corr}). 
    As a product of this strong gradient--size correlation, TNG50 also predicts a negative correlation between galaxy stellar mass and non-normalized gradient for redshifts $z \leq 2$ (Figure~\ref{fig:mstar}). 
    These results, in addition to the steepening of gradients with redshift, point to inside-out galaxy formation in TNG50.}
    \item{TNG50 accurately reproduces the gradient distribution observed in the local Universe, along with the observed correlation between galaxy stellar mass and gradient. 
    Moreover, TNG50 qualitatively recovers the characteristic normalized gradient observed at redshift $z \approx 0$. 
    Even at redshift $z \approx 0.5$, the TNG50 gradient distribution and gradient--stellar mass correlation all agree reasonably well with the measurements of \citet{Carton18}, the only study of this redshift range to date (Figures~\ref{fig:mstar}~and~\ref{fig:lspc}). }
    \item{At redshifts $z \gtrsim 1$, there exists tension between the predictions of TNG50 and observations. 
    TNG50 does not show the same preference for shallow-to-flat gradients observed by most large-sample studies at these redshifts.}
    This disagreement becomes more pronounced with redshift, as TNG50 gradients become more steep while observed gradients remain roughly constant. 
    Moreover, observations at these redshifts do not find evidence for the characteristic normalized gradient predicted by TNG50 (Figure~\ref{fig:gibsobs}). 
\end{enumerate}

While TNG50 agrees well with some observations of gradient distributions and correlations at redshifts $z \lesssim 1$, disagreements at redshifts $z \gtrsim 1$ (Sections~\ref{sec:dgradrobs}~and~\ref{sec:dmobs}) may indicate that the simulation does not accurately capture some phenomena (e.g. GMC-scale turbulence, particular characteristics of galactic winds) that work to flatten gradients at these redshifts. 
On the other hand, these disagreements may also be a result of observational biases and/or systematic errors in gradient measurements (Section~\ref{sec:dtension}). 
Future careful forward modeling (mocking) of these effects, including uncertain metallicity diagnostics and finite angular resolution/beam smoothing effects, applied to TNG50, can quantitatively demonstrate if this is the case. 
At the same time, observations from the next generation of telescopes (e.g. JWST/NIRSpec/NIRISS, ELTs) will eliminate several possible sources of systematics and make certain the state of metallicity gradients at redshifts $z \gtrsim 1$. 

\label{sec:conclusion}

\section*{Data availability}
The data that support the findings of this study are available on request from the corresponding author. Most of the data pertaining to the IllustrisTNG project is in fact already openly available on the IllustrisTNG website, \url{https://www.tng-project.org/data/}; those of the TNG50 simulation, in particular, are expected to be made publicly available within some months from this publication, at the same IllustrisTNG repository.

\section*{Acknowledgements}
ZSH acknowledges helpful correspondence with Ayan Acharyya and Mark Krumholz regarding metallicity gradient systematics.
ZSH thanks Sebastian F. S\'{a}nchez for assisting with low-redshift IFU survey comparisons and providing useful references.
MV acknowledges support through an MIT RSC award, a Kavli Research Investment Fund, NASA ATP grant NNX17AG29G, and NSF grants AST-1814053, AST-1814259 and AST-1909831.
FM acknowledges support through the Program ``Rita Levi Montalcini'' of the Italian MIUR.
The primary TNG simulations were realized with compute time granted by the Gauss Centre for Super-computing (GCS): TNG50 under GCS Large-Scale Project GCS-DWAR (2016; PIs Nelson/Pillepich), and TNG100 and TNG300 under GCS-ILLU (2014; PI Springel) on the GCS share of the supercomputer Hazel Hen at the High Performance Computing Center Stuttgart (HLRS). 
GCS is the alliance of the three national supercomputing centres HLRS (Universit\"{a}t Stuttgart), JSC (Forschungszentrum J\"{u}lich) and LRZ (Bayerische Akademie der Wissenschaften), funded by the German Federal Ministry of Education and Research (BMBF) and the German State Ministries for Research of Baden-W\"{u}rttenberg (MWK), Bayern (StMWFK) and Nordrhein-Westfalen (MIWF). 
Additional simulations for this paper were carried out on the Draco and Cobra supercomputers at the Max Planck Computing and Data Facility (MPCDF).








\appendix

\section{Resolution}

To test whether our TNG50-1 gradient results are well-converged, we also measure the metallicity gradients of lower-resolution TNG runs. Specifically, we measure the gradients of TNG50-2 (with $2 \times 1024^3$ resolution elements of mass $\sim 6.8 \times 10^5$) and TNG50-3 (with $2 \times 540^3$ resolution elements of mass $\sim 5.4 \times 10^6$). We generally follow the same TNG50-1 methodology when measuring the gradients of TNG50-2 and TNG50-3, although we are required to make a few minor changes to compensate for significantly reduced spatial resolution. We increase the TNG50-1 $M_{\mathrm{gas}} \geq 10^{9}$ sample cut to $M_{\mathrm{gas}} \geq 10^{9.5} \, M_{\odot}$ for TNG50-2, and to $M_{\mathrm{gas}} \geq 10^{10} \, M_{\odot}$ for TNG50-3 to ensure that each galaxy still posseses the $\sim 10^4$ gas cells required for robustly determining gradients. Moreover, we increase the pixel size of our metallicity maps used to derive metallicity profiles from 0.1 kpc for TNG50-1 to 0.25 kpc for TNG50-2 and to 0.5 kpc for TNG50-3. Finally, we relax the minimum particle requirements for accepting abundance and kinematic measurements.

\begin{figure}
\centerline{\vbox{\hbox{
\includegraphics[width=0.45\textwidth]{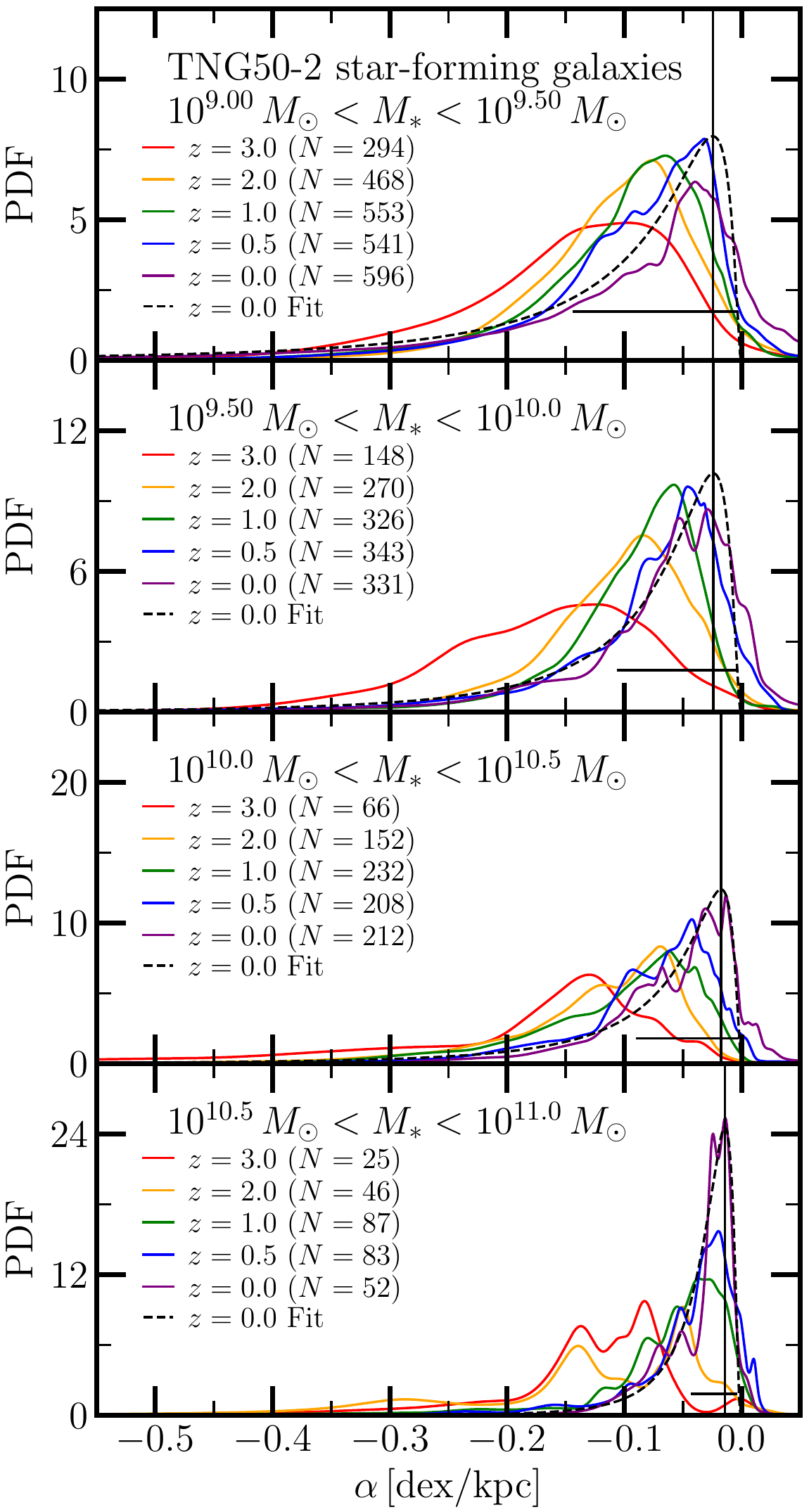}
}}}
\caption{
The metallicity gradient ($\alpha$) distributions of $10^9 \, M_{\odot} \leq M_* \leq 10^{11} \, M_{\odot}$ TNG50-2 star-forming galaxies, separated by stellar mass and redshift.
Each histogram is fit with a log-normal distribution -- for each mass bin, the redshift $z = 0$ fit is displayed as a black dashed curve. 
The peak of the redshift $z = 0$ log-normal fits are marked by black vertical lines, and the shortest spread around the peak that encloses 68\% of the distribution's probability is given as a horizontal black line.
The peak and spread of each log-normal fit are used in Table~\ref{tab:gradvza} to characterize each gradient distribution.}
\label{fig:dist1080}
\end{figure}

\begin{figure}
\centerline{\vbox{\hbox{
\includegraphics[width=0.45\textwidth]{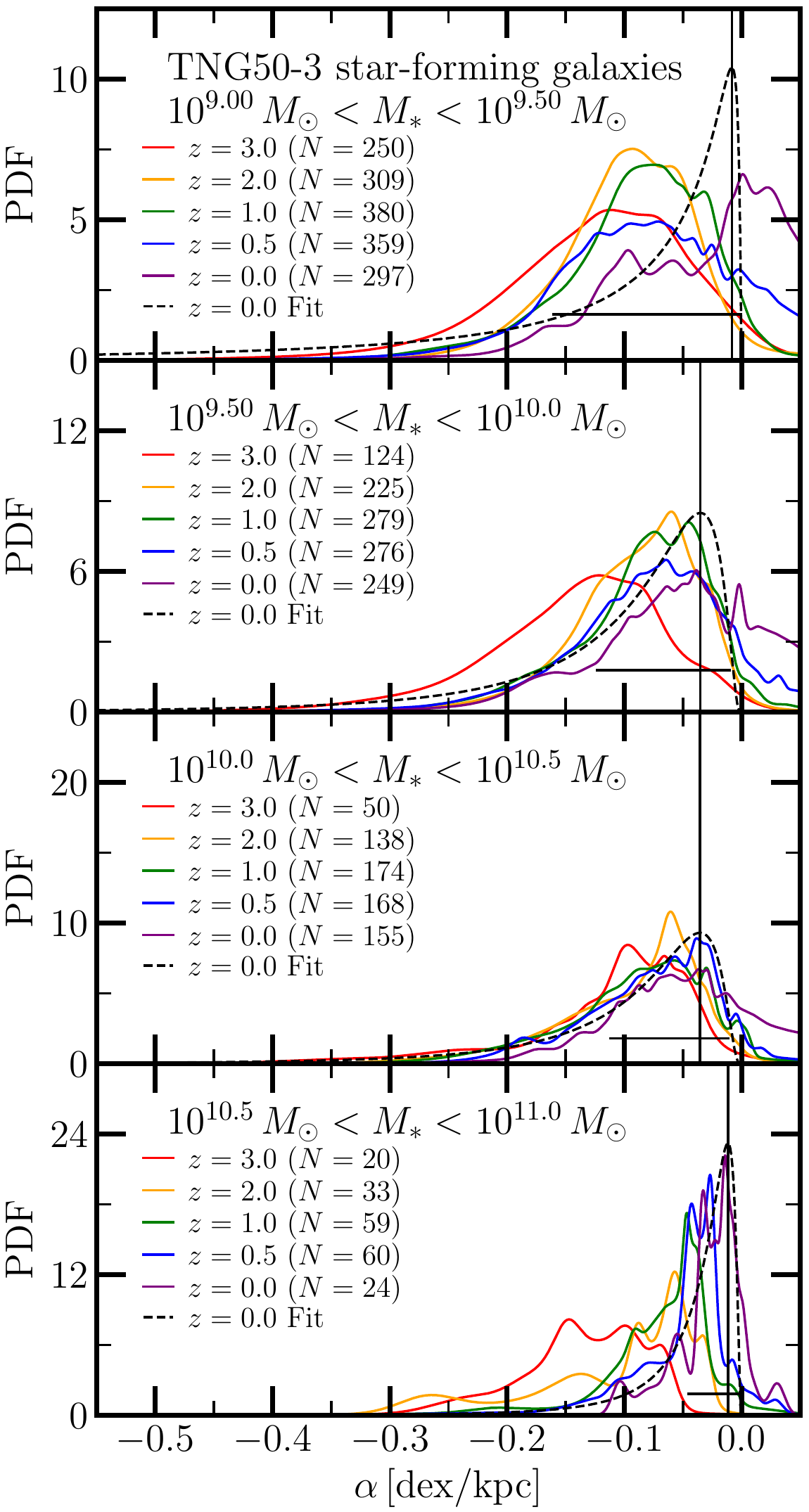}
}}}
\caption{The metallicity gradient ($\alpha$) distributions of $10^9 \, M_{\odot} \leq M_* \leq 10^{11} \, M_{\odot}$ TNG50-3 star-forming galaxies, separated by stellar mass and redshift.
Each histogram is fit with a log-normal distribution -- for each mass bin, the redshift $z = 0$ fit is displayed as a black dashed curve. 
The peak of the redshift $z = 0$ log-normal fits are marked by black vertical lines, and the shortest spread around the peak that encloses 68\% of the distribution's probability is given as a horizontal black line.
The peak and spread of each log-normal fit are used in Table~\ref{tab:gradvza} to characterize each gradient distribution.}
\label{fig:dist540}
\end{figure}

As we did for TNG50-1, we create distributions of the TNG50-2 and TNG50-3 gradients at each redshift (Figures~\ref{fig:dist1080}~and~\ref{fig:dist540}). We fit each of these distributions with log-normal distributions and quote the peak and spread of these distributions in Table~\ref{tab:gradvza}. This table also gives $\sigma_{\Delta \alpha}$, which is the offset between the TNG50-2/TNG50-3 gradients and the TNG50-1 gradient, normalized by the spread of the TNG50-1 distribution. We find $\sigma_{\Delta \alpha}$ to be generally low for TNG50-2, never growing beyond $|\sigma_{\Delta \alpha}| \sim 0.4$. On the other hand, we find more significant deviations for TNG50-3. In two cases, the TNG50-3 $\sigma_{\Delta \alpha}$ is greater than 1. Still, $\sigma_{\Delta \alpha}$ for TNG50-3 is invariably less than 0.5 otherwise. Thus, in most cases, the gradients of TNG50-2 and TNG50-3 are not significantly steeper or flatter than those of TNG50-1, although their distributions are significantly wider. Additionally, we find qualitatively similar correlations between TNG50-2/TNG50-3 gradients and galaxy stellar mass, size, and kinematics. Thus, we conclude that our TNG50-1 gradient results are reasonably well-converged.

\begin{table}                                                                    
    \centering
    \label{tab:gradvza}
    \caption{The redshift evolution of metallicity gradients in the $10^{9}\,M_{\odot} \leq M_* \leq 10^{11}\,M_{\odot}$ star-forming galaxies of TNG50-2 (top; with $2 \times 1080^3$ resolution elements) and TNG50-3 (bottom; with $2 \times 540^3$ resolution elements). Following our gradient methodology for TNG50-1, we separate galaxies into four stellar mass bins and fit the gradient distributions of these bins with log-normal distributions at each redshift. The quoted metallicity gradients and their uncertainties are the peak and spread of these log-normal fits.}
    \begin{tabular}{crrrrrrrr}                                                     
        \hline                                                                   
        $z$ & $M_*^{\mathrm{min}}$ & $M_*^{\mathrm{max}}$ & $N$ & $\alpha$ & $\sigma_{\Delta \alpha}$ \\ 
            & $\left[ \log \left( \frac{M_*}{M_{\odot}} \right) \right]$ & $\left[ \log \left( \frac{M_*}{M_{\odot}} \right) \right]$ & & $\left[ \mathrm{dex / kpc} \right]$ & \\ 
        \hline                                                                   
            $3.0$ & $ 9.0$ & $ 9.5$ & $294$ & $-0.090^{+0.051}_{-0.114}$ & $-0.154$ \\ [2pt] 
            $3.0$ & $ 9.5$ & $10.0$ & $148$ & $-0.120^{+0.059}_{-0.117}$ & $-0.338$ \\ [2pt] 
            $3.0$ & $10.0$ & $10.5$ & $ 66$ & $-0.117^{+0.056}_{-0.106}$ & $-0.101$ \\ [2pt] 
            $3.0$ & $10.5$ & $11.0$ & $ 25$ & $-0.097^{+0.038}_{-0.061}$ & $-0.255$ \\ [2pt] 
            $2.0$ & $ 9.0$ & $ 9.5$ & $468$ & $-0.073^{+0.038}_{-0.081}$ & $0.117$ \\ [2pt] 
            $2.0$ & $ 9.5$ & $10.0$ & $270$ & $-0.072^{+0.037}_{-0.077}$ & $0.074$ \\ [2pt] 
            $2.0$ & $10.0$ & $10.5$ & $152$ & $-0.076^{+0.038}_{-0.077}$ & $-0.164$ \\ [2pt] 
            $2.0$ & $10.5$ & $11.0$ & $ 46$ & $-0.050^{+0.036}_{-0.131}$ & $0.411$ \\ [2pt] 
            $1.0$ & $ 9.0$ & $ 9.5$ & $553$ & $-0.057^{+0.034}_{-0.083}$ & $0.216$ \\ [2pt] 
            $1.0$ & $ 9.5$ & $10.0$ & $326$ & $-0.058^{+0.030}_{-0.060}$ & $0.093$ \\ [2pt] 
            $1.0$ & $10.0$ & $10.5$ & $232$ & $-0.054^{+0.033}_{-0.087}$ & $0.101$ \\ [2pt] 
            $1.0$ & $10.5$ & $11.0$ & $ 87$ & $-0.024^{+0.017}_{-0.058}$ & $0.271$ \\ [2pt] 
            $0.5$ & $ 9.0$ & $ 9.5$ & $541$ & $-0.039^{+0.027}_{-0.089}$ & $0.196$ \\ [2pt] 
            $0.5$ & $ 9.5$ & $10.0$ & $343$ & $-0.036^{+0.024}_{-0.072}$ & $0.071$ \\ [2pt] 
            $0.5$ & $10.0$ & $10.5$ & $208$ & $-0.040^{+0.025}_{-0.069}$ & $-0.225$ \\ [2pt] 
            $0.5$ & $10.5$ & $11.0$ & $ 83$ & $-0.017^{+0.012}_{-0.046}$ & $0.212$ \\ [2pt] 
            $0.0$ & $ 9.0$ & $ 9.5$ & $596$ & $-0.024^{+0.020}_{-0.129}$ & $0.068$ \\ [2pt] 
            $0.0$ & $ 9.5$ & $10.0$ & $331$ & $-0.024^{+0.019}_{-0.082}$ & $-0.414$ \\ [2pt] 
            $0.0$ & $10.0$ & $10.5$ & $212$ & $-0.018^{+0.014}_{-0.072}$ & $-0.204$ \\ [2pt] 
            $0.0$ & $10.5$ & $11.0$ & $ 52$ & $-0.014^{+0.009}_{-0.028}$ & $0.087$ \\ [2pt] 
        \hline
        \hline
            $3.0$ & $ 9.0$ & $ 9.5$ & $250$ & $-0.082^{+0.046}_{-0.103}$ & $0.014$ \\ [2pt] 
            $3.0$ & $ 9.5$ & $10.0$ & $124$ & $-0.105^{+0.049}_{-0.091}$ & $-0.059$ \\ [2pt] 
            $3.0$ & $10.0$ & $10.5$ & $ 50$ & $-0.069^{+0.035}_{-0.071}$ & $0.450$ \\ [2pt] 
            $3.0$ & $10.5$ & $11.0$ & $ 20$ & $-0.108^{+0.041}_{-0.065}$ & $-0.483$ \\ [2pt] 
            $2.0$ & $ 9.0$ & $ 9.5$ & $309$ & $-0.070^{+0.035}_{-0.070}$ & $0.152$ \\ [2pt] 
            $2.0$ & $ 9.5$ & $10.0$ & $225$ & $-0.057^{+0.032}_{-0.071}$ & $0.242$ \\ [2pt] 
            $2.0$ & $10.0$ & $10.5$ & $138$ & $-0.054^{+0.030}_{-0.069}$ & $0.173$ \\ [2pt] 
            $2.0$ & $10.5$ & $11.0$ & $ 33$ & $-0.055^{+0.030}_{-0.068}$ & $0.340$ \\ [2pt] 
            $1.0$ & $ 9.0$ & $ 9.5$ & $380$ & $-0.052^{+0.032}_{-0.083}$ & $0.292$ \\ [2pt] 
            $1.0$ & $ 9.5$ & $10.0$ & $279$ & $-0.047^{+0.029}_{-0.074}$ & $0.260$ \\ [2pt] 
            $1.0$ & $10.0$ & $10.5$ & $174$ & $-0.052^{+0.031}_{-0.079}$ & $0.129$ \\ [2pt] 
            $1.0$ & $10.5$ & $11.0$ & $ 59$ & $-0.047^{+0.022}_{-0.040}$ & $-0.388$ \\ [2pt] 
            $0.5$ & $ 9.0$ & $ 9.5$ & $359$ & $-0.057^{+0.036}_{-0.099}$ & $-0.128$ \\ [2pt] 
            $0.5$ & $ 9.5$ & $10.0$ & $276$ & $-0.046^{+0.030}_{-0.088}$ & $-0.228$ \\ [2pt] 
            $0.5$ & $10.0$ & $10.5$ & $168$ & $-0.039^{+0.026}_{-0.076}$ & $-0.184$ \\ [2pt] 
            $0.5$ & $10.5$ & $11.0$ & $ 60$ & $-0.031^{+0.016}_{-0.031}$ & $-0.465$ \\ [2pt] 
            $0.0$ & $ 9.0$ & $ 9.5$ & $297$ & $-0.008^{+0.008}_{-0.151}$ & $0.263$ \\ [2pt] 
            $0.0$ & $ 9.5$ & $10.0$ & $249$ & $-0.035^{+0.025}_{-0.088}$ & $-1.193$ \\ [2pt] 
            $0.0$ & $10.0$ & $10.5$ & $155$ & $-0.035^{+0.024}_{-0.077}$ & $-1.879$ \\ [2pt] 
            $0.0$ & $10.5$ & $11.0$ & $ 24$ & $-0.012^{+0.009}_{-0.034}$ & $0.165$ \\ [2pt] 
        \hline
    \end{tabular}                                                                
\end{table}                                                                      

\bsp	
\label{lastpage}
\end{document}